\documentclass[11pt]{article}
\usepackage[margin=1.1in]{geometry}
\usepackage{amsmath,amssymb}
\usepackage{graphicx}
\usepackage{booktabs}
\usepackage{float}
\usepackage{array}
\usepackage[font=small,labelfont=bf]{caption}
\usepackage[colorlinks=true,linkcolor=black,citecolor=black,urlcolor=blue]{hyperref}

\newcommand{\tF}{t_{\mathrm{F}}}
\newcommand{\tN}{t_{\mathrm{N}}}
\newcommand{\tP}{t_{\mathrm{P}}}
\newcommand{\tS}{t^{\star}_{\mathrm{N}}}
\newcommand{\dd}{\mathrm{d}}
\DeclareRobustCommand{\S}{\ensuremath{\mathsection}}

\makeatletter\def\@fnsymbol#1{\ensuremath{\ifcase#1\or *\or \dagger\or \ddagger\fi}}\makeatother

\title{\textbf{Autocatalytic order controls bistability in a conservative kinetic model of relational event progression}\thanks{October 2026. Prepared as a replacement version of arXiv:2102.08826; version~1~\cite{boman2021arxiv} presents the earlier three-class model and remains available. This paper supersedes the formulation of~\cite{boman2025}. Relative to the August 2026 draft, this paper corrects the constant term of the asymptotic expansion \eqref{eq:asym}; states explicitly that temporal ordering is assumed rather than derived; adds the energy reading of the conserved budget as an interpretive postulate (\S4.4), the transition-state reading of the activated class (\S7), and the relational reading of renewal and a hypothesis about entropy (\S4.1, \S9); adds an explicit nondimensionalisation (\S2.4); and adds interior bistability with strictly irreversible activation and completion (\S6.7). All other results are unchanged and were re-checked with separately written code (Appendix~A).}}
\author{Bruce M. Boman\\[2pt]
\small Department of Mathematical Sciences, University of Delaware\\
\small Newark, DE 19716, USA}
\date{}

\begin{document}
\maketitle

\begin{abstract}
\noindent On a relational view, the Universe is not a container for events but the happening of events. A published kinetic model gave this dynamical form, sorting events as future, now, or past. Future events combine reversibly into now events; a now event meeting past events is activated before completing irreversibly into past events; past events are then renewed for future use. Past events thus catalyse their own production, and the network is closed, conserving a weighted event count. What such a network minimally requires to sustain distinct stable steady states of progression has not previously been established.

We give a complete mass-action and bifurcation analysis. Conservation fixes the stoichiometry uniquely. When a single past event participates in each activation, the past-event variable factors out and no fold can arise; progression ignites softly at a transcritical threshold, staying non-bistable even when closure is relaxed to vary the yield. When two past events participate cooperatively, the progressing branch folds: a past-rich attractor coexists with a past-poor state, and ignition demands a finite perturbation.

Two effects usually conflated are thus separated. Cooperative order creates memory; nonequilibrium asymmetry determines only how hard that memory is to undo, strict one-way latching being a singular limit of ordinary hysteresis whose recovery threshold scales as the square root of the reverse-flux scale. In a partially reversible network, in which only renewal is reversible and neither activation nor completion is ever reversed, both stable states lie in the interior, with a second fold in the budget at $E\simeq\tfrac12k_{-5}^{-2/3}$. Conservation does not prevent bistability. What governs it is the nonlinear order by which accumulated events influence their own continuation.

The work is entirely theoretical, and it does not derive time or its arrow: the relational ordering of events and the irreversibility of local transitions are assumed, and the paper establishes what that assumption requires and permits. As explicitly stated interpretive postulates, events are read as discrete parcels of energy and the conserved budget as conserved energy; the activated class is then the activated complex of transition-state theory, and the rate constants are ensemble propensities in Boltzmann's statistical sense. Around one turn of the cycle both the budget and the number of events are conserved, so at a running steady state the composition is stationary while energy is concentrated and dispersed step by step; whether this implies vanishing thermodynamic entropy production is left as a hypothesis. The postulates change no equation or result.
\end{abstract}

\noindent\textbf{Keywords:} relational time; event progression; autocatalysis; conservation law; saddle-node bifurcation; bistability; transition-state theory

\section{Introduction}

This paper develops a kinetic model of relational event progression in which future, now, activated-now, and past events are represented by an autocatalytic reaction network~\cite{boman2025}. The question we pursue is structural: what does a closed, conservative event network minimally require in order to sustain two distinct stable steady states of progression, rather than a single one? We answer it by analysing the full mass-action system with an exact conserved budget, complete steady-state solutions, Jacobian stability analysis on the invariant leaf, and bifurcation geometry.

The answer is that the participation of past events must be cooperative. When a single past event participates in each activation, the network supports a transcritical exchange of stability---a genuine ignition threshold in the event budget---but not two coexisting stable steady states. Second-order participation, in which two past events take part, produces the fold, the coexistence of attractors, finite-amplitude ignition and latching. Since the earlier analysis of this network proposed two stable equilibria and hysteretic switching for the first-order mechanism~\cite{boman2025}, our result also revises that expectation, and we set out below where the two analyses diverge.

The motivation for treating the system as closed is ontological rather than merely convenient. On the relational view the paper inherits---and in Rovelli's formulation, that the world is not a collection of things to which events happen but is itself the happening of events~\cite{rovelli2018,rovelli2002}---there is no exterior from which event activity could be supplied or into which it could drain. Modelling the event system without external source or sink terms is therefore not an idealisation adopted for tractability but what the ontology suggests: if the totality of what there is consists in the happening of events, the system has no outside from which to draw. That argument establishes autonomy, however, and not conservation of any particular quantity. A system with no exterior may still create, annihilate, branch or merge events internally, and need not conserve a selected weighted count. Conservation of the positive linear event-weight functional introduced in \S4.2 is accordingly an additional postulate: it does not follow from the absence of an exterior, and its role is to define the closed subclass of networks analysed here.

The correction is specific and, we think, constructive. The published criterion for bistability was the simultaneous vanishing of two partial balances, $\dd(\tF+\tN)/\dd s = \dd(\tS+\tP)/\dd s = 0$. That condition identifies a steady state in which two sub-balances are separately stationary; it does not establish two complete stable equilibria separated by a basin boundary, which is what bistability requires and what the Jacobian settles. When the full system is analysed, the first-order mechanism is found to possess exactly two steady-state branches, which meet and exchange stability at a critical budget. The dynamical phenomenon it exhibits is soft, reversible, threshold ignition---which is itself a substantive result about event progression---but not two coexisting attractors and not hysteresis.

The obstruction turns out to be sharply localised. Within the fixed network topology studied here it is caused neither by the conservation law nor by the choice of decomposition yield, but by the autocatalytic step being first order in the past-event variable, and it is removed by making that step cooperative. We hold the topology fixed throughout and vary only the kinetic order, so what is shown is that changing the topology is unnecessary---not that topology plays no role in broader classes of event network.

\paragraph{What is assumed, and what is shown.} The paper does not derive time, and it does not derive the arrow of time. The relational classification of events as future, now and past, and the irreversibility of the local transitions between them, are inputs to the model. What the paper establishes is conditional on those inputs: given a closed, conserved network of relational events with an irreversible local ordering, it determines exactly what is required for the progression to sustain two distinct, persistent, history-dependent stable steady states rather than one. The contribution is therefore a statement about the dynamics that a relational ordering supports, not an account of where the ordering comes from.

\paragraph{What the model is, and what the events are.} The work is entirely theoretical: the network is a proposed model, no data are fitted to it, and every result is a property of the stated equations. The mathematics proves that a weighted budget $E$ is exactly conserved; it does not, and cannot, prove what $E$ physically is. We therefore state the physical reading separately, as an interpretive postulate (\S4.4): events are discrete parcels of energy, the class weights are energy content per event, and $E$ is conserved energy. On that reading the network is a kinetic scheme of the kind transition-state theory describes---a reversible pre-equilibrium feeding a high-energy activated complex that commits irreversibly to products (\S7)---and the rate constants are what they are in that theory: ensemble propensities summarising an innumerable population of microscopic interactions, in the statistical sense introduced by Boltzmann (\S3.3). The postulate changes no equation, no number and no result; it states what the equations are taken to describe.

\subsection{Relation to existing work}

Second-order autocatalysis producing a saddle-node fold is not itself a new phenomenon. The Schl\"ogl model~\cite{schlogl1972}, with autocatalysis second order in the autocatalyst, is the standard minimal example of a chemical fold, and cubic autocatalysis plays the same role in the Gray--Scott scheme~\cite{grayscott1984}. Chemical reaction network theory supplies complementary structural results: the deficiency theorems~\cite{hornjackson1972,feinberg1987} and injectivity-based criteria~\cite{craciun2005} bound multistationarity from network structure. The control coefficients and summation theorem used in \S5.4 are standard metabolic control analysis~\cite{kacser1973,heinrich1974}. Fast-intermediate elimination of the kind used in \S7 is classical singular perturbation~\cite{tikhonov1952,segel1989}, and bistable fronts in such systems are well studied~\cite{fife1977}. The kinetic architecture invoked in \S7---a pre-equilibrium feeding an activated complex---is that of transition-state theory~\cite{eyring1935,evanspolanyi1935,hanggi1990}.

The contribution here is therefore not the existence of folds under cooperative autocatalysis. It is the diagnosis of the obstruction in the previously proposed conserved event network~\cite{boman2025}, the correction of its bistability claim, and the identification of the minimal event-interaction order required within that framework. Two features distinguish the setting from the classical models: those models are open, holding reservoir species fixed, which supplies driving and removes any conservation law, whereas here the dynamics are confined to a leaf of fixed budget; and the negative result proved in \S5.1---that the factorisation persists for every decomposition yield $n$---appears not to have been stated for this class.

The model has a lineage. Its first form~\cite{boman2021arxiv} had three classes---future, now and past---with now taken as the transition state, self-renewal steps generating future and now events, and no conservation law, so that its abundances grew without bound; a successor~\cite{boman2020} and the published model~\cite{boman2025} introduced past-event autocatalysis and the activated class. The present paper closes the network, which the first form did not, and the analysis below shows why closure matters: it is what fixes the stoichiometry and makes a comparison of stable steady states meaningful.

The relational reading of events on which the state variables rest is developed separately~\cite{boman2026tdil,boman2026rc}; \S2 imports it rather than re-deriving it. The present paper is the kinetic arm of that programme and inherits its ontology unchanged.

\subsection{Candidate physical settings}

The network is synthetic and we fit no data to it. It is nonetheless worth recording which classes of real system share its two defining features---a conserved total pool and self-amplification of order higher than one in the amplifier---since Appendix~C shows that any realisation must be driven rather than at equilibrium.

Protein aggregation with secondary nucleation is the closest structural analogue known to us: total protein mass is conserved on the aggregation timescale, monomer is converted into fibril mass, and secondary nucleation makes growth superlinear in existing fibril mass, producing a lag phase, sharp autocatalytic takeoff, and seed-dependent onset~\cite{knowles2009,cohen2013}. Glycolytic oscillators are a second candidate, the adenine nucleotide pool being conserved while phosphofructokinase is activated by its own product~\cite{higgins1964,selkov1968,goldbeter1996}. Both are driven systems with a conserved subpool, which is precisely the configuration Appendix~C identifies as necessary. We have not attempted to fit either, and offer them as orientation rather than as claimed instances.

\section{The relational events being counted}

\subsection{From relational events to a kinetic ensemble}

Following the companion framework~\cite{boman2026rc}, and consonant with relational formulations in which measurements are correlations between dynamical quantities rather than readings against an external parameter~\cite{rovelli2002,rovelli2018}, a physical event is a distinguishable transition or interaction in a physical system---an atomic transition, an emission, a collision, a change of configuration---and progression is the accumulation of distinguishable cycles, transitions, phases or counts in one process, always assessed relative to another. Progression is a comparative notion; no absolute rate is presupposed.

We treat events as discrete parcels of energy. The notion of discrete energy units is as old as Planck~\cite{planck1901}; we use the term \emph{parcel} rather than \emph{quantum} to mark that the model concerns the progression of events, not the quantization of a field or oscillator. How much energy each class of event carries, and what that implies for the conserved budget, is set out in \S4.4.

The labels future, now and past accordingly describe the relational status of such events with respect to a specified reference process or event ensemble. They are not absolute designations: the same event carries different status with respect to different reference processes. The analysis is written at a single reference: all four abundances are components of one state vector defined against one reference ensemble. The relational character of the labels enters the model at one point only, the renewal step R5, which \S4.1 reads as the single-reference representation of exchange with other references.

The position taken on time is the one physics took on heat. Heat was once modelled as a substance, caloric, that flowed from hot bodies to cold ones; it is now understood as the interaction of molecules, with no substance flowing, and it is known through a measurement, temperature. Time is treated here in the same way: not as a substance that flows, but as the interaction of events, known through a measurement---the comparison of one process with another, which is what a clock performs~\cite{boman2026rc}.

It follows, and is worth stating explicitly because the notation invites the opposite reading, that $\tF, \tN, \tP, \tS$ are not quantities of a temporal substance in different states. Nothing in this model flows, and no entity called time is transported between the classes. What the variables count is physical events---transitions and interactions---sorted by their relational status. The conserved budget $E$ introduced in \S4.2 is accordingly a budget of events, and on the reading of \S4.4 a budget of their energy; it is not a budget of time.

The variables $\tF$, $\tN$ and $\tP$ are therefore not coordinates in spacetime but coarse-grained non-negative state variables representing the abundance of event activity assigned to each relational class, relative to a fixed reference ensemble. We call these \emph{event-class abundances}.

\paragraph{Two senses of ``state''.} The word is used at two levels, which must not be confused. At the level of the individual event, $\tF$, $\tN$, $\tS$ and $\tP$ are \emph{energy states of events}: the condition, and the energy content (\S4.4), of an event at a given stage of progression. The now and activated-now states are transient---an event passes through them---and the activated-now state is the least persistent of all. At the level of the network, a \emph{steady state} is a composition of the whole ensemble that no longer changes with progression, and it may be stable or unstable. The bifurcation results concern steady states of the network: the past-rich and past-poor steady states are stable, the separatrix is a saddle steady state, unstable but with a stable manifold, and none of them is an energy state of an individual event. No spatial denominator is intended and they are not densities in the per-volume sense; where a particular class is meant we write, for instance, \emph{past-event abundance} for the abundance of the past class. The fourth variable $\tS$ represents activated now events---now events engaged in an interaction with past events and committed to the progression, but not yet completed. Section~7 identifies this class with the activated complex of transition-state theory.

Two things follow that matter for what can be claimed. The analysis is carried out at one reference; beyond the lumped renewal step of \S4.1 it says nothing about how the labelling transforms between reference processes. And because the variables are coarse-grained populations rather than individual events, all conclusions are statements about ensemble dynamics at the chosen descriptive scale.

\subsection{The progression variable}

The independent variable is $s$, and it is not time. Time enters only through the dependent variables $\tF,\tN,\tP,\tS$, which count events sorted by relational status; were time also the independent variable, the model would presuppose the flow it sets out to describe, and the circularity would be fatal.

Here $s$ measures change of any kind: a parameter that advances whenever anything changes, without commitment as to what changes---displacement in space, a configurational rearrangement, a state transition, an interaction. This generality is what the relational premise demands. If an event is determinable only in relation to other events, the parameter against which events are counted cannot be a privileged external coordinate; it must be constituted by change itself.

Formally, $s$ is a monotone path parameter on trajectories in the coarse-grained event-state space. We state plainly what is not yet supplied: the model gives no unique physical prescription for constructing $s$ from heterogeneous microscopic changes. It does not specify whether one collision, one atomic transition, one configurational change and one displacement increment contribute equally to $\dd s$, nor how an observer would calibrate them against one another. Supplying such a prescription would make $s$ an operational observable; until then it is a formal ordering parameter, and the results below are correspondingly interpreted as statements about oriented trajectories, equilibria and bifurcations rather than about rates.

\paragraph{Relation to earlier usage, and to the companion framework.} In the published model the same symbol appears in the elementary event relation $E=\dd t/\dd s$ with $s$ a spatial displacement~\cite{boman2025}. The companion paper~\cite{boman2026rc} subsequently demotes that quantity to a motivating heuristic, on three grounds: it is trajectory-bound and undefined for a stationary process, where $\dd s = 0$; its three-dimensional extension $\nabla t$ is a synchronisation convention and can be gauged away; and its dimensions do not match the dimensionless clock comparisons through which gravitational time dilation is measured.

Widening $s$ to count change of any kind meets the first objection directly, and this is the main reason for the wider definition. An atomic transition occurring in place---the paradigm event of the companion framework---produces no displacement but is itself a distinguishable change, so it advances $s$ even though it advances no spatial coordinate. On the wider reading, $s$ is precisely progression in the companion paper's sense: the accumulation of distinguishable transitions or counts. The second objection does not arise here, since $s$ is an ordering parameter for an ensemble rather than a field assigned over space, so there is nothing to resynchronise. The third is addressed below under units.

One distinction should be kept clear. The companion framework's fundamental object is the event field $\vec{E} = \nabla\chi$, built from clock-rate ratios and defined over space for static configurations. That is a different object from $s$, serving a different purpose: $\vec{E}$ describes how progression varies with position, whereas $s$ counts progression itself. The present paper uses only the latter and makes no use of the event field.

\paragraph{Units and the circularity question.} Let the event-class variables carry abundance units $D$ and let $s$ carry units $S$ of change. Then from $r_i$ having units $S^{-1}D$: the first-order constants $k_2,k_4,k_5$ carry $S^{-1}$; the second-order constant $k_1$ carries $S^{-1}D^{-1}$; and the autocatalytic constant $k_3$, appearing in $r_3 = k_3\tN\tP^m$, carries $S^{-1}D^{-m}$---that is, $S^{-1}D^{-1}$ for $m=1$ and $S^{-1}D^{-2}$ for $m=2$. Numerical values quoted with ``all $k_i=1$'' are therefore understood in units where $D$ and $S$ are both set to unity, and the variables are treated as dimensionless abundances throughout.

By construction $S$ denotes a unit of formal progression rather than of physical time, and the rate constants are accordingly expressed per unit $s$. The dimensional bookkeeping records that choice rather than establishing it: what rules out circularity is the definition of $s$ in this section, not the absence of a time symbol from the constants. With $s$ so defined, the model does not use physical time as the parameter generating temporal classification.

\paragraph{Steady state need not be static.} A steady state is a condition in which the event-class abundances are unchanging with respect to further change. On a running branch this does not mean nothing happens: the composition is stationary while the progression $\tF\to\tN\to\tS\to\tP$ continues at constant turnover. The past-poor equilibrium is different in kind---an extinguished state with zero cycle turnover---so ``steady'' covers both a running and a halted condition, and which is meant will be stated where it matters.

\paragraph{What depends on the parametrisation.} Under a positive constant rescaling of the unit of progression, rate constants rescale correspondingly and only the rate at which trajectories are traversed changes; all steady states, branches and bifurcation locations are unaffected. More generally, a strictly monotone reparametrisation $s\mapsto\sigma(s)$ preserves the oriented state-space trajectories, since $\dd x/\dd\sigma = f(x)/\sigma'(s)$, but it need not preserve an autonomous parametrisation: for non-constant $\sigma'$ the transformed system is explicitly $\sigma$-dependent. Only the oriented trajectories, equilibria and bifurcations are interpreted here; their traversal rate depends on the chosen parametrisation, and growth rates and threshold-crossing rates are quoted per unit $s$ as defined.

\subsection{Notation and units}

Table~\ref{tab:notation} collects the variables and parameters used below in terms of the two base units introduced above: $D$ for event-class abundance and $S$ for progression. Table~\ref{tab:abbrev} lists the abbreviations.

\begin{table}[t]
\centering\small
\begin{tabular}{lp{7.2cm}ll}
\toprule
Symbol & Meaning & Units & Defined\\
\midrule
$s$ & progression parameter; counts change of any kind & $S$ & \S2.2\\
$\tF$ & future-event abundance & $D$ & \S2.1\\
$\tN$ & now-event abundance & $D$ & \S2.1\\
$\tS$ & activated-now abundance (activated complex, \S7) & $D$ & \S2.1\\
$\tP$ & past-event abundance & $D$ & \S2.1\\
$w_X$ & event weight of class $X$; $(w_F,w_N,w_P,w_\star) = (1,2,1,m{+}2)$; energy per event on the reading of \S4.4 & --- & \S4.2\\
$E$ & conserved event budget, $\mathbf{w}^\top x$; conserved energy on the reading of \S4.4 & $D$ & \S4.2\\
$m$ & order of past-event participation in R3 & --- & \S4.1\\
$n$ & completion yield of R4; closure forces $n=m+2$ & --- & \S4.2\\
$k_1$ & forward constant, $2\tF\to\tN$ & $S^{-1}D^{-1}$ & \S4.3\\
$k_2$ & reverse constant, $\tN\to2\tF$ & $S^{-1}$ & \S4.3\\
$k_3$ & autocatalytic constant, $\tN + m\tP\to\tS$ & $S^{-1}D^{-m}$ & \S4.3\\
$k_4$ & completion constant, $\tS\to n\tP$ & $S^{-1}$ & \S4.3\\
$k_5$ & renewal constant, $\tP\to\tF$ & $S^{-1}$ & \S4.3\\
$\epsilon$ & dimensionless reversibility scale, $k_{-i}=\epsilon\kappa_{-i}$ & --- & App.~B\\
$\kappa_{-3}$ & reverse constant of R3$'$ at unit $\epsilon$ & $S^{-1}$ & App.~B\\
$\kappa_{-4}$ & reverse constant of R4$'$ at unit $\epsilon$ & $S^{-1}D^{-3}$ & App.~B\\
$\kappa_{-5}$ & reverse constant of R5 at unit $\epsilon$ & $S^{-1}$ & App.~B\\
$K_1$ & equilibrium constant $k_1/k_2$ & $D^{-1}$ & App.~C\\
$K_3$ & equilibrium constant $k_3/k_{-3}$ & $D^{-m}$ & App.~C\\
$K_4$ & equilibrium constant $k_4/k_{-4}$ & $D^{n-1}$ & App.~C\\
$K_5$ & equilibrium constant $k_5/k_{-5}$ & --- & App.~C\\
$r_i$ & reaction flux of step $i$ & $S^{-1}D$ & \S4.3\\
$J$ & steady-state cycle turnover, $J\equiv r_3$ & $S^{-1}D$ & \S5.4\\
$C_i$ & flux control coefficient, $\partial\ln J/\partial\ln k_i$ & --- & \S5.4\\
$\lambda$ & eigenvalue on the invariant leaf & $S^{-1}$ & \S6.2\\
$E_c$ & critical budget of the transcritical bifurcation & $D$ & \S5.2\\
$E_{\mathrm{fold}}$ & budget at the fold & $D$ & \S6.2\\
$V_X$ & postulated potential energy per event of class $X$ & --- & \S4.4\\
$E_a$, $E_a^{\mathrm{app}}$ & activation energy of the activation step; apparent activation energy (postulated increments) & --- & \S4.4\\
$k_{-5}$ & reverse-renewal constant, $\tF\to\tP$ & $S^{-1}$ & \S6.7\\
$E_{\mathrm{low}}$, $E_{\mathrm{up}}$ & lower and upper fold budgets of the partially reversible network & $D$ & \S6.7\\
$t_{\mathrm{P,fold}}$ & past-event abundance at the fold & $D$ & \S6.2\\
$t^{*}_{\mathrm{P}}$ & ignition threshold for a stated family of initial conditions & $D$ & \S6.3\\
$\pi_{\to}$ & forward-commitment probability of the activated complex & --- & \S7\\
\bottomrule
\end{tabular}
\caption{Variables and parameters, with units in terms of event-class abundance $D$ and progression $S$. Note that the reverse constant of the first pair is written $k_2$ rather than $k_{-1}$, following the numbering of the scheme. The Wegscheider combination $K_1K_3K_4K_5^2$ of Appendix~C is dimensionless, as it must be, for $m=2$, $n=4$.}
\label{tab:notation}
\end{table}

\begin{table}[H]
\centering\small
\begin{tabular}{lp{10.2cm}}
\toprule
Abbreviation & Meaning\\
\midrule
R1, R2 & condensation $2\tF\to\tN$ and its reverse $\tN\to2\tF$ (\S4.1)\\
R3, R4 & activation $\tN+m\tP\to\tS$ and completion $\tS\to n\tP$, general order $m$ (\S4.1)\\
R3$'$, R4$'$ & cooperative activation $\tN+2\tP\to\tS$ and completion $\tS\to4\tP$, $m=2$ (\S6.1)\\
R5 & renewal $\tP\to\tF$ (\S4.1)\\
TST & transition-state theory (\S7)\\
AI & artificial intelligence (AI-assistance statement)\\
Radau IIA & implicit Runge--Kutta integration method used for trajectories (App.~A)\\
LSODA & adaptive solver for stiff and non-stiff differential equations, used for verification (App.~A)\\
RK45, DOP853, BDF & explicit Runge--Kutta (orders 4/5 and 8) and backward-differentiation integration methods (App.~A)\\
CO & carbon monoxide (\S7.1)\\
\bottomrule
\end{tabular}
\caption{Abbreviations used in the text.}
\label{tab:abbrev}
\end{table}

\subsection{Nondimensionalisation, and two scales kept apart}

The statement ``all $k_i=1$'' is made precise as follows. Let $A_0$ be a characteristic event-abundance scale (units $D$) and $S_0$ a characteristic progression scale (units $S$), and define $x_X = t_X/A_0$ and $u = s/S_0$. For the cooperative model the dimensionless constants are
\[
a_1 = k_1A_0S_0,\quad a_2 = k_2S_0,\quad a_3 = k_3A_0^{2}S_0,\quad a_4 = k_4S_0,\quad a_5 = k_5S_0,
\]
and the conserved budget becomes $e = x_F + 2x_N + x_P + 4x_\star = E/A_0$. The balances in $x$ and $u$ have the form of \eqref{eq:F}--\eqref{eq:S} with $k_i\to a_i$, so every result quoted at unit constants is a result at $a_i=1$. Positive constant rescaling of $A_0$ and $S_0$ leaves the phase portrait, the equilibria and their stability unchanged.

Two scales must not be confused. $A_0$ is an abundance scale; it is not an energy scale. Under the energy reading of \S4.4 each event of class $X$ carries $w_X\varepsilon_0$, where $\varepsilon_0$ is an energy per unit weight, so the physical energy is postulated to be $\varepsilon_0A_0e$. The weights $(1,2,1,4)$ are dimensionless relative energy contents; $\varepsilon_0$ is not determined by the model, and supplying it is the unit map left open in \S9.

\section{Modelling assumptions}

Because this paper sits on an inherited conceptual framework and adds new kinetic hypotheses to it, we separate the two explicitly, together with what is demonstrated and what is interpretation.

\paragraph{A. Inherited conceptual framework~\cite{boman2025,boman2026rc}.} Temporal measurement does not require an independently existing entity that flows: a clock is a physical process, and duration is a comparison between processes. Physical events are distinguishable transitions or interactions. Progression is comparative, presupposing no absolute rate. Temporal status---future, now, past---is relational to a reference process. Notably, the companion framework leaves open how causal order is encoded and how the apparent arrow of time emerges from asymmetries among event relations, listing irreversible interactions among the candidate sources and assigning the question to a complete theory~\cite{boman2026rc}. The present paper does not answer that question; it assumes an irreversible local ordering and asks what follows.

\paragraph{B. New kinetic hypotheses introduced here.} The relational classification of events and the irreversibility of selected local transitions are assumed, not derived. Event classes admit continuous population variables. Interactions are approximated by mass-action laws. The network is well mixed at the chosen descriptive scale. A positive linear event budget is conserved. Selected transitions are irreversible. Past-event participation may be first or second order.

\paragraph{C. Results demonstrated below.} First-order factorisation permits only two branches, which exchange stability transcritically, for every decomposition yield. Second-order dependence produces a saddle-node fold. Past-poor and past-rich states coexist on the same invariant leaf. Ignition is finite amplitude. Irreversible kinetics produce one-way latching. The fast-intermediate limit yields a reciprocal symmetry, conditional on $k_2=k_5$.

\paragraph{D. Interpretive postulates and proposals, not established.} That events are discrete parcels of energy, the class weights are energy content per event, and the conserved budget is conserved energy (\S4.4); the quantitative map from budget units to physical energy is not supplied. That the activated class is the activated complex of transition-state theory, and the rate constants are ensemble propensities in the statistical sense (\S3.3, \S7). That renewal R5 represents exchange between references, an event past for one reference being future for another (\S4.1); and the hypothesis that entropy, increasing locally at each completion, has zero global change without appeal to a low-entropy initial condition (\S9). That the model may represent self-sustaining event progression; that past-rich and past-poor states may correspond to distinct regimes of progression. None of these readings alters an equation, a number or a result in category~C.

\subsection{Why mass action}

Mass-action kinetics do not follow from the relational definition of an event. They are adopted as a minimal ensemble closure: the frequency of a transition involving one now-class event and $m$ past-class events is taken proportional to $\tN\tP^m$. This assumption is not derived from the relational ontology; it defines the dynamical model to be tested. It presumes that the ensemble is well mixed at the descriptive scale, so that encounter frequencies factorise into products of class abundances.

The two cases examined are $m=1$, in which a single past event participates in each activation---the mechanism of the published model---and $m=2$, in which two past events participate cooperatively. Higher $m$ is a straightforward extension and is not pursued, since $m=2$ is already sufficient.

\subsection{Why closure}

The relational ontology motivates modelling the total event system without external source or sink terms: if the Universe is the happening of events rather than a container in which events occur~\cite{rovelli2018}, there is no exterior to supply or absorb event activity. It does not, however, dictate which internal quantity is preserved, since a system without an exterior may still create or destroy event weight internally. We therefore additionally impose conservation of a positive linear event-weight functional as a closure postulate. The absence of an exterior does not derive this particular invariant; rather, the invariant defines the closed subclass analysed here.

Closure also does three pieces of technical work: it prevents event activity from being created or destroyed by unmodelled external influence; it bounds the dynamics, so trajectories cannot escape to infinity; and it makes comparison of attractors meaningful, since two states are comparable only when they carry the same budget. Given the selected reaction scheme and the normalisation $w_F=1$, the requirement fixes the admissible weights and decomposition yield uniquely, as shown next.

\subsection{Rate constants as ensemble propensities}

Boltzmann's step was to give up tracking each molecule's position and momentum and to treat macroscopic thermodynamic quantities as statistical averages over an immense number of microscopic occurrences~\cite{boltzmann1877}. Reaction-rate theory took the same step for chemical change: a rate constant does not describe any single molecular encounter; it summarises how often, per unit of the governing parameter, an encounter drawn from a population of jostling molecules with a distribution of energies carries the system across the activation barrier~\cite{eyring1935,evanspolanyi1935,hanggi1990}.

The constants $k_1,\dots,k_5$ are read in the same spirit, with one distinction kept explicit. A mass-action coefficient is not itself a probability: it carries order-dependent units (Table~\ref{tab:notation}), and the flux $r_i$ it defines is a rate per unit progression. What the statistical reading supplies is its meaning. Each event interaction is not followed individually; $k_i$ is the propensity coefficient---the ensemble-averaged rate per unit progression per combination of participating events---with which an interaction of the stated type proceeds through the stated transition, and the probability that a given encounter succeeds within an increment $\dd s$ is proportional to $k_i\,\dd s$. The mass-action fluxes $r_i$ are the corresponding ensemble-averaged rates. This is what licenses the coarse-grained, population-level description of \S2.1, and it is why every conclusion of the paper is a statement about ensemble dynamics rather than about individual events.

The reading is interpretive and limited. The model contains no temperature, no partition functions and no energy distribution over individual parcels, so the values of $k_i$ are not computed from first principles; they are free parameters, and all numerical results are reported at stated values. What the statistical reading supplies is the meaning of those parameters, not their magnitudes.

\section{The conserved network}

\subsection{Scheme}

\begin{align*}
&\text{R1: } \tF+\tF \xrightarrow{k_1} \tN &&\text{R2: } \tN \xrightarrow{k_2} \tF+\tF\\
&\text{R3: } \tN + m\,\tP \xrightarrow{k_3} \tS &&\text{R4: } \tS \xrightarrow{k_4} n\,\tP\\
&\text{R5: } \tP \xrightarrow{k_5} \tF
\end{align*}

R1/R2 is the reversible pair in which event activity fluctuates between future and now status; R3 is the autocatalytic step in which past events catalyse entry of a now event into the activated class; R4 is the irreversible completion to past status; R5 is renewal: activity classified as completed, or past, re-enters the pool capable of participating in future transitions.

R5 carries the relational content of the model and is read as follows. Because temporal status is relational, an event that is past with respect to one reference process is future with respect to others; the progression is perpetual not because energy is re-concentrated at any one reference, but because the past of one reference is the future of another~\cite{boman2025}. In the single-reference model analysed here R5 is a phenomenological, lumped renewal term: the first-order step $\tP\to\tF$ at rate $k_5$. Two things must be kept distinct. A change in relational classification---the same event counted as past by one reference and future by another---is not by itself a physical transfer of energy, and the model does not say how the one gives rise to the other. The mapping between reference classifications, energy exchange and the value of $k_5$ awaits an explicit multi-reference network and is deferred with the spatial extension (\S9). The proceedings account~\cite{boman2025} motivates the reading; it does not supply the rate law. The reading is an interpretive postulate (category~D); R5 is used below only as the stated mass-action step.

\subsection{The budget fixes the stoichiometry}

Seek weights $w_X>0$ with $\sum_X w_X\nu_{iX}=0$ for every step. R1 gives $w_N = 2w_F$; R5 gives $w_P=w_F$; R3 gives $w_\star = w_N + m\,w_P$; R4 gives $w_\star = n\,w_P$. With $w_F=1$ these are simultaneously satisfiable only when
\begin{equation}
n = m+2,\label{eq:closure}
\end{equation}
giving weights $(w_F,w_N,w_P,w_\star) = (1,2,1,m+2)$ and conserved budget
\begin{equation}
E = \tF + 2\tN + \tP + (m+2)\,\tS.\label{eq:budget}
\end{equation}
For $m=1$ this requires $\tS\to3\tP$; for $m=2$ it requires $\tS\to4\tP$. The published scheme uses $\tS\to2\tP$ with $m=1$, for which the two determinations of $w_\star$ conflict (3 from R3 against 2 from R4) and no positive linear conserved functional of this form exists: each turn of the loop destroys one unit of budget. Closure therefore selects the yield, and we adopt \eqref{eq:closure} throughout.

Each now event is assembled from two future events and so carries two units; each activated event carries the weight of the now event plus the participating past events. $E$ is the total budget of elementary event parcels in the system in whatever bound form. Note that the raw class count $\tF+\tN+\tP+\tS$ is not conserved and is not intended to be: R1 reduces it, R4 increases it.

\subsection{Balances}

With mass-action fluxes $r_1=k_1\tF^2$, $r_2=k_2\tN$, $r_3=k_3\tN\tP^m$, $r_4=k_4\tS$, $r_5=k_5\tP$, the balances $\dd X/\dd s = \sum_i \nu_{iX}r_i$ read
\begin{align}
\frac{\dd\tF}{\dd s} &= 2k_2\tN + k_5\tP - 2k_1\tF^2 \label{eq:F}\\
\frac{\dd\tN}{\dd s} &= k_1\tF^2 - k_2\tN - k_3\tN\tP^m \label{eq:N}\\
\frac{\dd\tP}{\dd s} &= (m+2)\,k_4\tS - m\,k_3\tN\tP^m - k_5\tP \label{eq:P}\\
\frac{\dd\tS}{\dd s} &= k_3\tN\tP^m - k_4\tS \label{eq:S}
\end{align}
The multiplier $m$ on the autocatalytic flux appears in the $\tP$ balance \eqref{eq:P}, because each activation consumes $m$ past events; the $\tS$ balance \eqref{eq:S} carries no such multiplier, since one activated event is produced per occurrence.

The multiplier 2 on $k_1\tF^2$ in \eqref{eq:F} is likewise stoichiometric and independent of reaction order. R1 is second order, expressed by the $\tF^2$ dependence of $r_1$; the multiplier 2 expresses that each occurrence consumes two future events. The two are separate: R2 is first order yet also carries a factor 2, because each occurrence liberates two future events. Omitting the multiplier on the forward term while retaining it on the reverse gives $\dd E/\dd s = +k_1\tF^2 \neq 0$; integrating that unbalanced system from $E(0)=9$ yields $E=379$ at $s=10$ and divergence thereafter, with no steady state of any kind. With \eqref{eq:F} as written, substitution into \eqref{eq:budget} gives $\dd E/\dd s = 0$ identically.

Because $E$ is conserved, phase space foliates into invariant three-dimensional leaves. All stability statements below are relative to the leaf; the full Jacobian carries a neutral eigenvalue normal to it, associated with the invariant, which is excluded throughout.

\subsection{The energy reading of the weights}

The analysis so far establishes that a unique positive weighting is conserved; it is silent on what that weighting is. We now state the physical reading adopted in the programme, as an interpretive postulate.

\paragraph{Postulate (energy reading).} Each event is a discrete parcel of energy, and the weight $w_X$ of a class is the energy carried per event of that class. A future event carries one unit; a now event carries two, because it is the condensed pair; an activated event carries $m+2$ units---four in the cooperative model---the pair together with the past events gathered into it; a past event carries one. The conserved budget \eqref{eq:budget} is then conservation of energy.

\begin{figure}[t]
\centering
\includegraphics[width=0.95\textwidth]{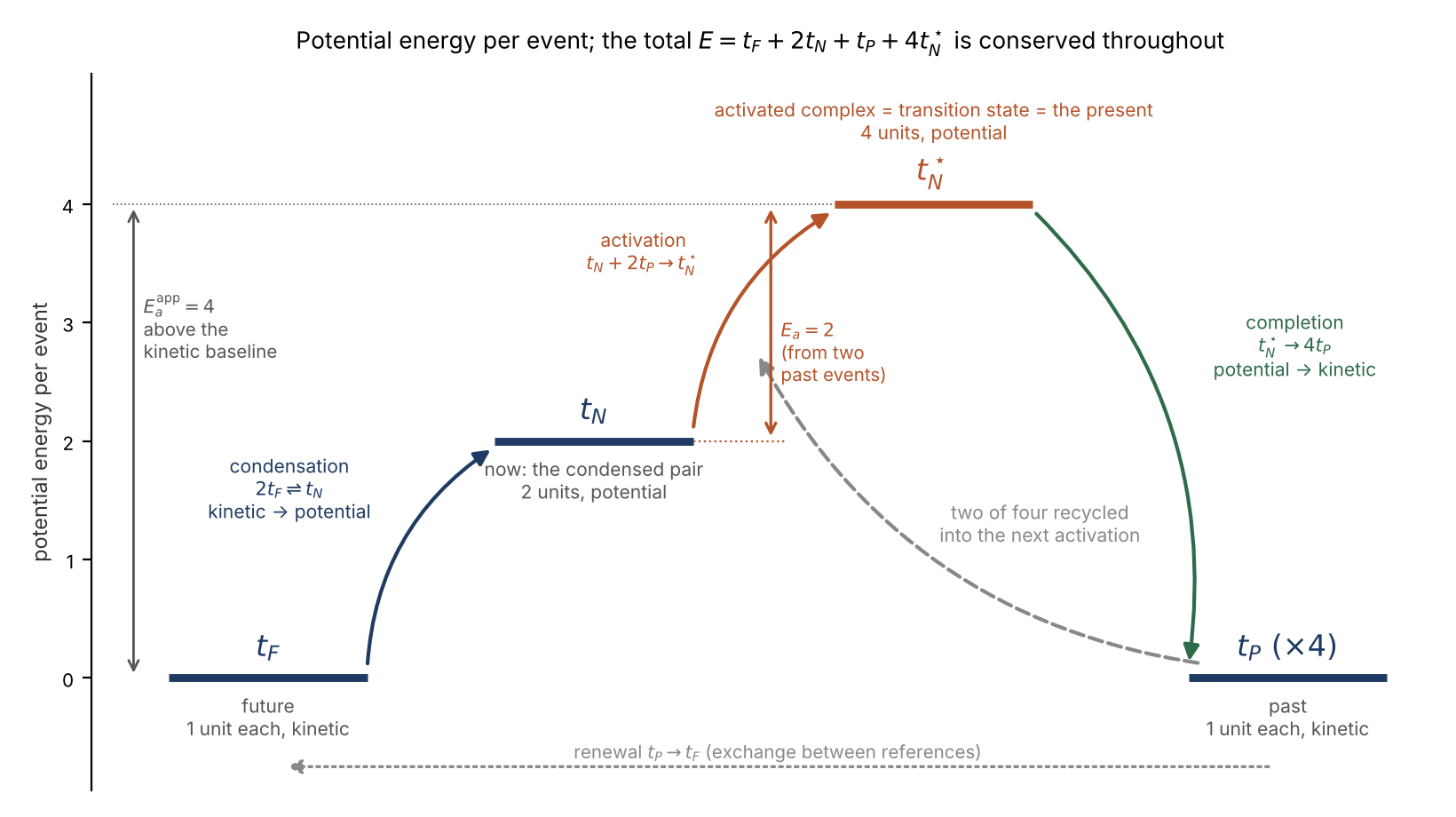}
\caption{The cooperative cycle ($m=2$) read as energy per event. Condensation concentrates two one-unit parcels into a two-unit now event (reversibly); activation gathers two past events into a four-unit activated complex at the top of the energy ladder; completion disperses the four units into four past events, two of which supply the next activation; renewal returns past to future status. Plotted is potential energy per event \eqref{eq:V}: future and past events carry their unit of energy as kinetic energy (potential zero), now and activated events as potential energy; the total $E$ is unchanged at every step. The activation step raises the potential energy by $E_a=2$ units \eqref{eq:Ea}, supplied by the kinetic energy of two past events; the peak lies $E_a^{\mathrm{app}}=4$ units above the kinetic baseline \eqref{eq:Eapp}.}
\label{fig:energy}
\end{figure}

The event classes are assigned dimensionless relative energy contents, not universal absolute energies. To read the activation step in the sense of transition-state theory one further distinction is needed, and it is made here as part of the postulate: the form in which an event carries its energy. Future and past events carry their energy as \emph{kinetic} energy---the energy of free, dispersed events available to interact. Condensation of two future events into a now event converts their kinetic energy into \emph{potential} energy: a now event holds its two units as potential energy. Activation raises the potential energy further, gathering the kinetic energy of $m$ past events into the activated event, which holds $m+2$ units as potential energy, the highest of any event. Completion releases that potential energy as the kinetic energy of $m+2$ past events. Writing $V_X$ for the potential energy per event,
\begin{equation}
V_F = 0,\qquad V_N = 2,\qquad V_\star = m+2,\qquad V_P = 0.\label{eq:V}
\end{equation}
Total energy is conserved at every step, as it is in every chemical reaction; what changes is its form. That is precisely the transition-state picture: the activation energy is not a gain in total energy but kinetic energy of the reactants converted into potential energy of the activated complex, and reactants that cannot supply it never reach the complex~\cite{eyring1935,evanspolanyi1935}.

Two activation energies follow. Both are postulated relative potential-energy increments, defined by the kinetic/potential division just stated; they are the analogues of activation energies under that postulate, not Eyring activation free energies derived from a reaction coordinate, and the activated class at the reference parameters is a populated intermediate rather than an ideal dividing surface (\S7). The activation energy of the activation step, measured from its reactant state,
\begin{equation}
E_a = V_\star - V_N = m,\label{eq:Ea}
\end{equation}
is two units in the cooperative model, supplied by the kinetic energy of the two participating past events. Because activation is preceded by the condensation pre-equilibrium, the activation energy of the progression as a whole, measured from the kinetic baseline of future events, is the sum of the pre-equilibrium step and the activation step,
\begin{equation}
E_a^{\mathrm{app}} = (V_N - 2V_F) + E_a = V_\star - 2V_F = m + 2,\label{eq:Eapp}
\end{equation}
four units in the cooperative model. This is the familiar result for a mechanism with a pre-equilibrium, in which the apparent activation energy is the sum of the pre-equilibrium energy change and the activation energy of the subsequent step; $E_a^{\mathrm{app}}$ is the height of the potential-energy peak above the baseline. Equations~\eqref{eq:Ea}--\eqref{eq:Eapp} link the energy reading to the paper's central result: the cooperative order $m=2$ that the bifurcation analysis requires for bistability is the number that sets the activation energy of the step at two units and the peak at four.

What the model does not supply should be stated with equal precision. The division of each event's energy into kinetic and potential is part of the postulate, not something the equations determine: the conserved budget counts total energy only. There is no temperature and no distribution of energy over events, so $k_3$ is not computed from $E_a$ through a Boltzmann factor, and the quadratic activation flux expresses cooperative encounter dependence; deriving the rate coefficient from $E_a$, or an activation free energy $\Delta G^{\ddagger}$ from an explicit microscopic model, is a separate problem not addressed here.

Three further features of the reading should be stated precisely.

First, it derives the stoichiometry rather than licensing changes to it. On the energy reading the multiplier 2 in \eqref{eq:F} is not bookkeeping: condensation removes two units' worth of future because two units' worth of future is what a now event is made of. The reading therefore applies to the reaction fluxes $r_i$, which remain hypotheses open to revision, and not to the stoichiometric coefficients, which are the conservation law. Any reinterpretation that altered a coefficient would break $\mathbf{w}^\top N = 0$, and the unbalanced variant of \S4.3 shows what follows.

Second, the transitions concentrate and disperse energy; they do not create it. Condensation places two units that lived in two events into one; activation gathers four units into a single event; completion distributes them into four kinetic past events (Fig.~\ref{fig:energy}). Per-event energy climbs $1\to2\to4$ and returns to 1, while the total is untouched. This is the sense in which progression is a sequence of energy transitions---between concentrated and dispersed forms---within an exactly conserved total.

Third, the reading is a postulate and is labelled as one. Conservation of the budget is a theorem about the network; conservation of energy is a law of physics; the postulate identifies the first with the second, and nothing in the mathematics forces that identification. The equations establish that $E$ is conserved; they do not establish that $E$ is energy. The quantitative map from budget units $D$ to physical energy is not supplied here and remains open. Nothing numerical in this paper depends on the postulate, and every result in category~C of \S3 holds whether or not it is adopted.

\section{First-order autocatalysis cannot produce a fold in this network class}

\subsection{The factorisation}

Set $m=1$. Setting \eqref{eq:S} to zero gives $k_3\tN\tP = k_4\tS$; substituting into \eqref{eq:P} at steady state,
\begin{equation}
n\,k_3\tN\tP - k_3\tN\tP - k_5\tP = 0 \;\Longrightarrow\; \tP\bigl[(n-1)k_3\tN - k_5\bigr] = 0.\label{eq:factor}
\end{equation}
The past-event variable appears as a common factor to the first power, leaving a bracket linear in $\tN$. Exactly two branches follow: a past-poor branch with $\tP=\tS=0$ and $k_1\tF^2 = k_2\tN$, existing at every budget; and a past-rich branch with $\tN = k_5/[(n-1)k_3]$, independent of the budget.

\paragraph{Relaxing closure to test the yield.} Conservation of the event budget fixes $n=3$ when $m=1$, so within the closed model the yield is not free. To test whether the absence of a fold is an accident of that particular value, we temporarily relax closure and allow non-conservative variants with arbitrary $n\ge2$. The steady-state condition \eqref{eq:factor} is unchanged in form, so varying the yield shifts the threshold $t^*_{\mathrm N} = k_5/[(n-1)k_3]$ but never creates a third branch. The obstruction therefore lies in the first-order dependence on $\tP$, not in the closure-selected value $n=3$---and it is not repaired by leaving the conservative subclass.

\paragraph{Scope of the claim.} What is proved is that for this topology, mass-action closure and first-order dependence on $\tP$, only the two stated branches occur, for every yield $n\ge2$. Within the present smooth steady-state geometry, two stable steady states on a leaf require an intervening unstable state and hence at least three steady states, so bistability is unavailable. What is not proved is that every conceivable first-order event-progression network is incapable of bistability; a different topology, or a closure other than mass action, is not covered.

\subsection{What occurs instead: soft ignition}

\begin{figure}[t]
\centering
\includegraphics[width=0.72\textwidth]{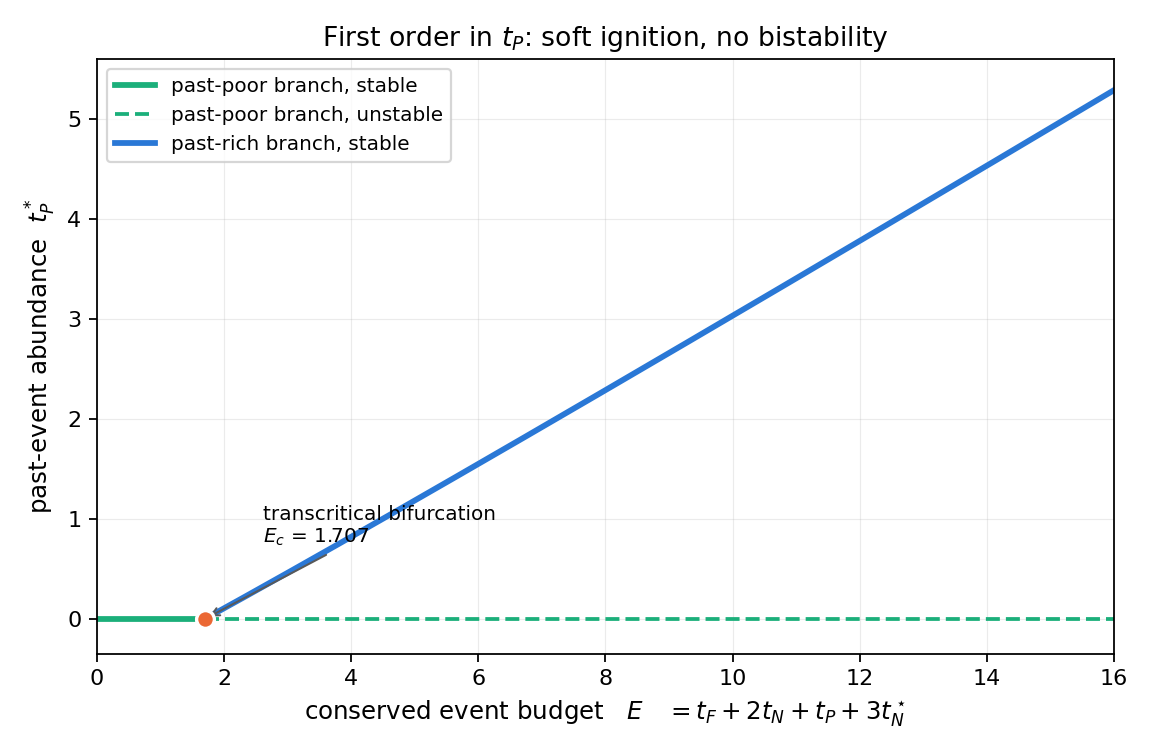}
\caption{Steady states of the first-order model against the conserved budget. Solid: stable on the fixed-$E$ leaf; dashed: unstable. The past-poor branch loses stability to the past-rich branch at $E_c=1.707$ through a transcritical bifurcation. The past-rich branch rises monotonically and does not fold back, so two stable steady states never coexist.}
\label{fig:first}
\end{figure}

Linearising about the past-poor branch, the $(\tP,\tS)$ block is
\[
J = \begin{pmatrix} -k_3\tN - k_5 & 3k_4\\ k_3\tN & -k_4\end{pmatrix},\qquad \det J = k_4\,(k_5 - 2k_3\tN),
\]
so the past-poor state is stable on the leaf while $\tN < k_5/(2k_3)$ and becomes a saddle above. Since $\tN$ on that branch increases monotonically with $E$, stability is lost at
\begin{equation}
E_c = \sqrt{\frac{k_2k_5}{2k_1k_3}} + \frac{k_5}{k_3} = 1.7071\quad\text{(unit constants)},\label{eq:Ec}
\end{equation}
where the past-rich branch emerges with $\tP=0$ and rises monotonically (Fig.~\ref{fig:first}). The branches exchange stability: a transcritical bifurcation, not a fold. No budget supports two stable steady states.

Ignition here is \emph{soft}: above $E_c$ an arbitrarily small population of past events grows, because the past-poor state is linearly unstable. The threshold lies in the budget, not in the size of the perturbation, and the transition is continuous and reversible.

\subsection{Why: the past does not accelerate its own production}

On the past-rich branch $\tN$ is pinned at $k_5/[(n-1)k_3]$, so the loop turnover is
\begin{equation}
J = k_3\tN\tP = \frac{k_5}{n-1}\,\tP,\label{eq:Jlin}
\end{equation}
exactly linear in the past-event abundance. The per-capita growth rate of $\tP$ is therefore set by $\tN$ and is independent of $\tP$: accumulating past events raises the absolute rate of past-event production but not its own specific rate. Accumulation instead depletes $\tN$, and growth ceases when $\tN$ reaches the pinned value. For the past to drive its own production in a way that can latch, more past must make past accumulate faster \emph{per unit past}; with one past event consumed per activation, the past merely keeps pace with itself.

\subsection{Distribution of control}

All steady-state fluxes in a cycle are rigidly related---at $E=9$ with unit constants the net flux through the $\tF\rightleftharpoons\tN$ pair, $r_3$ and $r_4$ all equal 1.3295, and $r_5 = 2.6590$---so rate limitation is assessed by control coefficients $C_i = \partial\ln J/\partial\ln k_i$. Throughout, $J$ denotes the steady-state cycle turnover, taken as the flux through the autocatalytic step, $J\equiv r_3 = k_3\tN\tP^m$, evaluated at the past-rich steady state; since $r_3=r_4$ there, the choice of representative step is immaterial.

\begin{table}[H]
\centering\small
\begin{tabular}{llr}
\toprule
Step & & $C_i$\\
\midrule
R4 & $\tS\to3\tP$ & $+0.5587$\\
R5 & $\tP\to\tF$ & $+0.2065$\\
R3 & $\tN+\tP\to\tS$ & $+0.1660$\\
R1 & $2\tF\to\tN$ & $+0.0947$\\
R2 & $\tN\to2\tF$ & $-0.0259$\\
\midrule
sum & & $+1.0000$\\
\bottomrule
\end{tabular}
\caption{Control coefficients for the first-order model at $E=9$, all $k_i=1$; the sum satisfies the summation theorem.}
\end{table}

Completion of the activated event carries the majority of control, being the sole source of past events, which are simultaneously the co-substrate of R3 and the autocatalyst. R2 carries small negative control, as expected of a step returning now events to future status. R1 carries least, consistent with operating close to its own equilibrium. Control is distributed: there is a dominant step but no strictly rate-limiting one.

\subsection{Quality of the future--now quasi-equilibrium}

The model treats R1/R2 as a fast quasi-equilibrium. Its quality is the ratio of net to gross forward flux, which should be small. With unit constants it is 0.73: the pair is drained faster than it re-equilibrates and is not a quasi-equilibrium at all. Scaling $k_1,k_2$ by $\lambda$ gives 0.22 at $\lambda=10$, 0.028 at $\lambda=100$, 0.0096 at $\lambda=300$. In the stiff limit the control coefficients of R1 and R2 become equal and opposite ($+0.049$, $-0.048$), summing to zero: the pair sets the state but contributes nothing to the rate. The condition to state is $k_1,k_2\gg k_3\tP$, whereupon $\tN\approx(k_1/k_2)\tF^2$.

\section{Cooperative second-order participation}

\subsection{Formulation}

Set $m=2$, so that activation requires two past events:
\[
\text{R3}':\ \tN + 2\tP \xrightarrow{k_3} \tS,\qquad \text{R4}':\ \tS\xrightarrow{k_4} 4\tP,
\]
with budget $E = \tF + 2\tN + \tP + 4\tS$ by \eqref{eq:closure}. The cause-and-effect relation of the published model is unchanged---past events increase the propensity of a now event to enter the activated class---but the influence is now cooperative, growing with the square of past-event abundance.

\subsection{The fold}

\begin{figure}[t]
\centering
\includegraphics[width=0.72\textwidth]{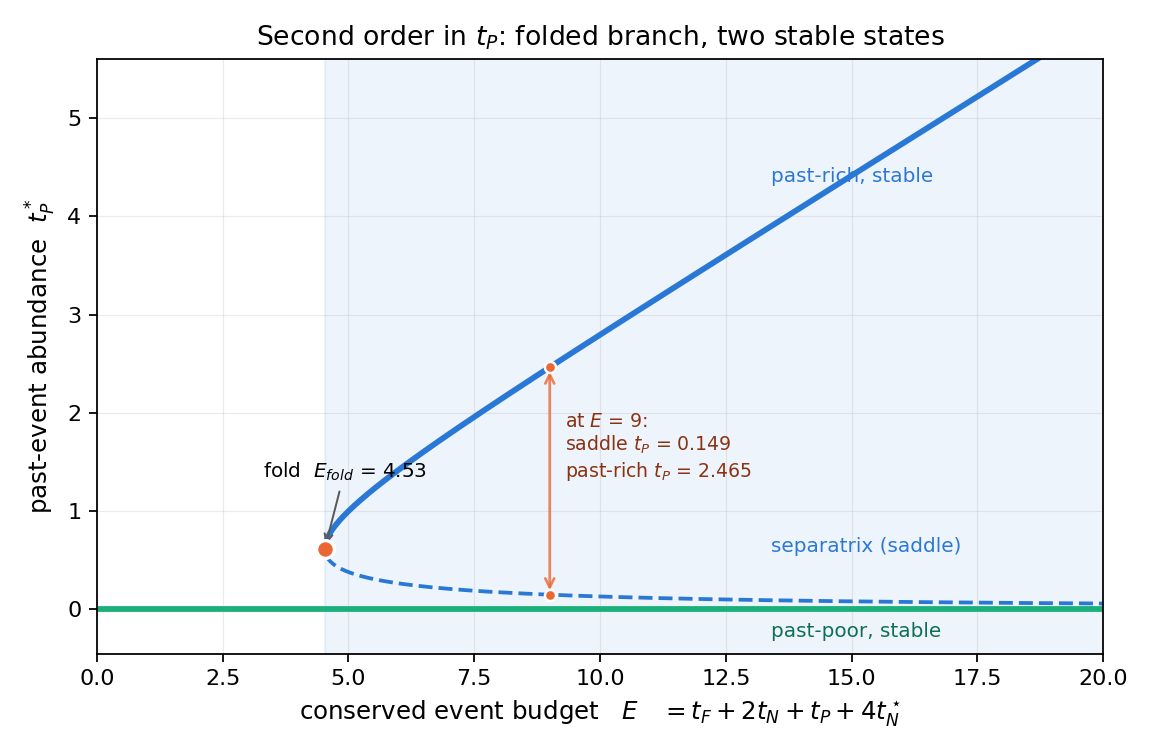}
\caption{Steady states of the cooperative model. The past-rich branch folds at $E_{\mathrm{fold}}=4.53$; above the fold the past-rich state, the saddle, and the past-poor state coexist on the same leaf. At $E=9$ the saddle lies at $\tP=0.149$ and the past-rich state at $\tP=2.465$.}
\label{fig:fold}
\end{figure}

The condition analogous to \eqref{eq:factor} is
\begin{equation}
\tP\,(2k_3\tN\tP - k_5) = 0,\label{eq:factor2}
\end{equation}
a \emph{quadratic} bracket. The past-rich branch satisfies $\tN\tP = k_5/(2k_3)$---a hyperbola rather than a pinned constant. Parameterising by $\tP$,
\begin{gather}
\tN = \frac{k_5}{2k_3\tP},\qquad \tS = \frac{k_5\tP}{2k_4},\qquad \tF = \sqrt{\frac{\tN\,(k_2+k_3\tP^2)}{k_1}},\nonumber\\
E(\tP) = \tF + \frac{k_5}{k_3\tP} + \tP + \frac{2k_5\tP}{k_4}.\label{eq:Ebranch}
\end{gather}
The term $k_5/(k_3\tP)$ diverges as $\tP\to0$ while the last grows linearly, so $E(\tP)$ has an interior minimum: the fold, at $t_{\mathrm{P,fold}}=0.618$ and $E_{\mathrm{fold}}=4.5295$ for unit constants. Above $E_{\mathrm{fold}}$, \eqref{eq:Ebranch} has two roots---an upper stable steady state and a lower saddle---which with the past-poor state give three coexisting steady states on the leaf (Fig.~\ref{fig:fold}).

Crucially, the past-poor state is now stable at \emph{every} budget: linearising about $\tP=\tS=0$, the $\tP^2$ term vanishes to second order and the block is triangular with eigenvalues $-k_5$, $-k_4$, both negative regardless of $\tN$. There is no critical budget at which it destabilises, which is what allows genuine coexistence rather than succession. Note that this state lies on the boundary of the positive orthant rather than in its interior.

\paragraph{The physically admissible variant.} That boundary location matters, and the qualification is stated here rather than deferred. On the relational reading of \S2, a configuration in which a class is empty is a degeneracy of the labelling, not a state of the network. The physically admissible form of the result is given in \S6.7: a network in which activation and completion remain strictly one-way and only renewal is given a small reverse rate $k_{-5}$. There every steady state is strictly interior, all four classes populated: at $E=9$ and $k_{-5}=10^{-3}$ the three states carry $\tP=1.9\times10^{-3}$ (past-sparse, stable), $0.147$ (saddle) and $2.47$ (past-rich, stable). The strictly irreversible network analysed here is the $k_{-5}\to0$ limit: it gives the bifurcation structure in its sharpest analytical form, while the network of \S6.7 supplies the realisation that the ontology admits without reversing the commitment of any event. A fully weakly reversible network, in which activation and completion are reversed as well, is treated in Appendix~B as a comparison. Section~9 returns to the distinction.

\begin{table}[H]
\centering\small
\begin{tabular}{lcccc}
\toprule
 & $\tF$ & $\tN$ & $\tP$ & $\tS$\\
\midrule
past-poor (stable) & 1.8860 & 3.5570 & 0 & 0\\
separatrix (saddle) & 1.8508 & 3.3508 & 0.1492 & 0.0746\\
past-rich (stable) & 1.1981 & 0.2028 & 2.4654 & 1.2327\\
\bottomrule
\end{tabular}
\end{table}

Both stable steady states are steady with respect to change, not static: the past-rich state sustains turnover 1.2327 per unit change indefinitely, the past-poor state none. The now-event abundance differs by a factor of 17.5 between them. Eigenvalues on the leaf, excluding the neutral direction normal to it, are

\begin{table}[H]
\centering\small
\begin{tabular}{ll}
\toprule
state & leaf eigenvalues\\
\midrule
past-poor (stable) & $-1,\ -1,\ -8.544$\\
saddle & $+0.221,\ -4.234,\ -8.412$\\
past-rich (stable) & $-2.580\pm1.236\,i,\ -10.710$\\
\bottomrule
\end{tabular}
\end{table}

\noindent The single positive eigenvalue at the saddle confirms a codimension-one separatrix within the leaf, and the complex pair indicates damped oscillatory approach to the past-rich state.

\subsection{Finite-amplitude ignition}

\begin{figure}[t]
\centering
\includegraphics[width=0.72\textwidth]{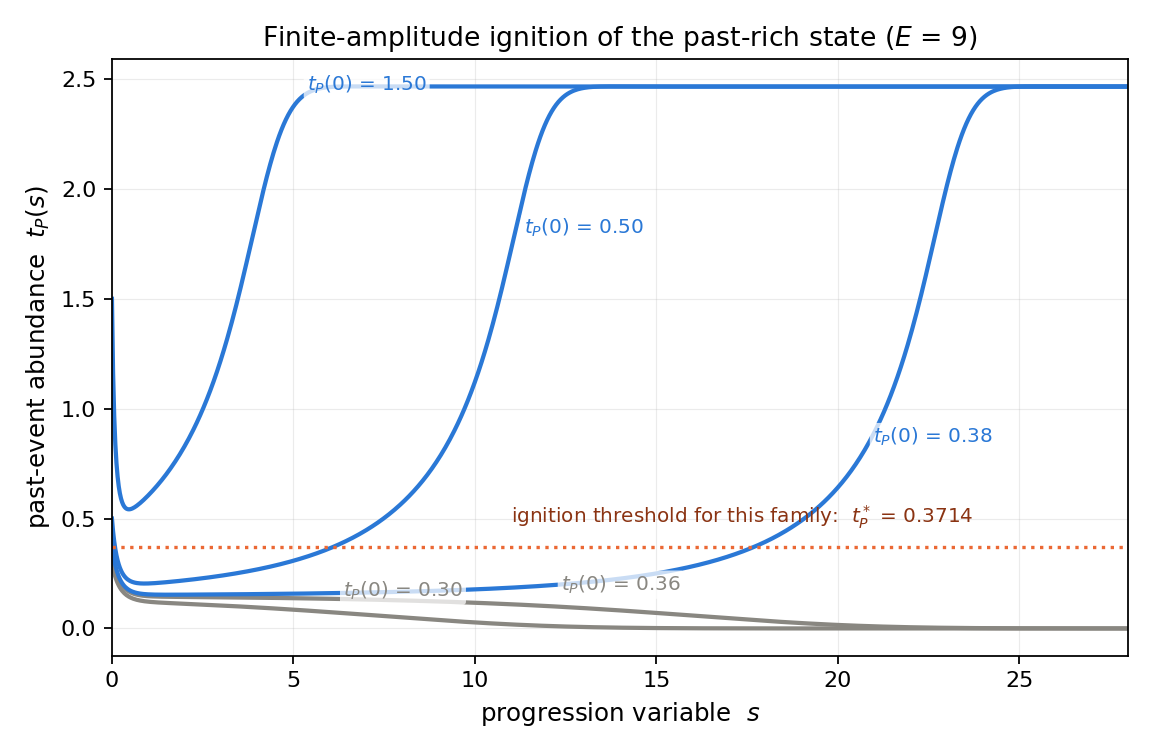}
\caption{Ignition trajectories at $E=9$, labelled by initial past-event abundance $\tP(0)$. Grey: seeds below threshold decay to the past-poor state. Blue: seeds above threshold ignite to the past-rich state. All trajectories dip initially, since each activation consumes two past events before the cooperative loop can replace them; the labels mark the curve, not its starting height. The threshold \emph{for this family of initial conditions} is $t^*_{\mathrm P} = 0.3714$.}
\label{fig:ignition}
\end{figure}

Because the past-poor state is stable at every budget, ignition requires a finite perturbation. Figure~\ref{fig:ignition} integrates initial conditions at $E=9$ with $\tS(0)=0$, $\tP(0)$ as labelled, and $\tF,\tN$ carrying the remaining budget at the past-poor ratio. The threshold for this family is $t^*_{\mathrm P} = 0.3714$.

This is not the saddle's own coordinate, 0.1492. The separatrix is the saddle's stable manifold, a two-dimensional surface within the leaf, not a surface of constant $\tP$; the measured threshold is its intersection with the chosen line of initial conditions. Different preparations at the same budget give different apparent thresholds, so any threshold statement must specify the preparation.

\subsection{One-way latching}

We distinguish three notions. \emph{Bistability} is coexistence of two stable steady states on a leaf, established above. \emph{Closed-loop hysteresis} is history dependence in which an up-and-down parameter sweep crosses two folds. \emph{One-way latching} is history dependence with a single switching event. The present system exhibits the third.

\begin{figure}[t]
\centering
\includegraphics[width=0.72\textwidth]{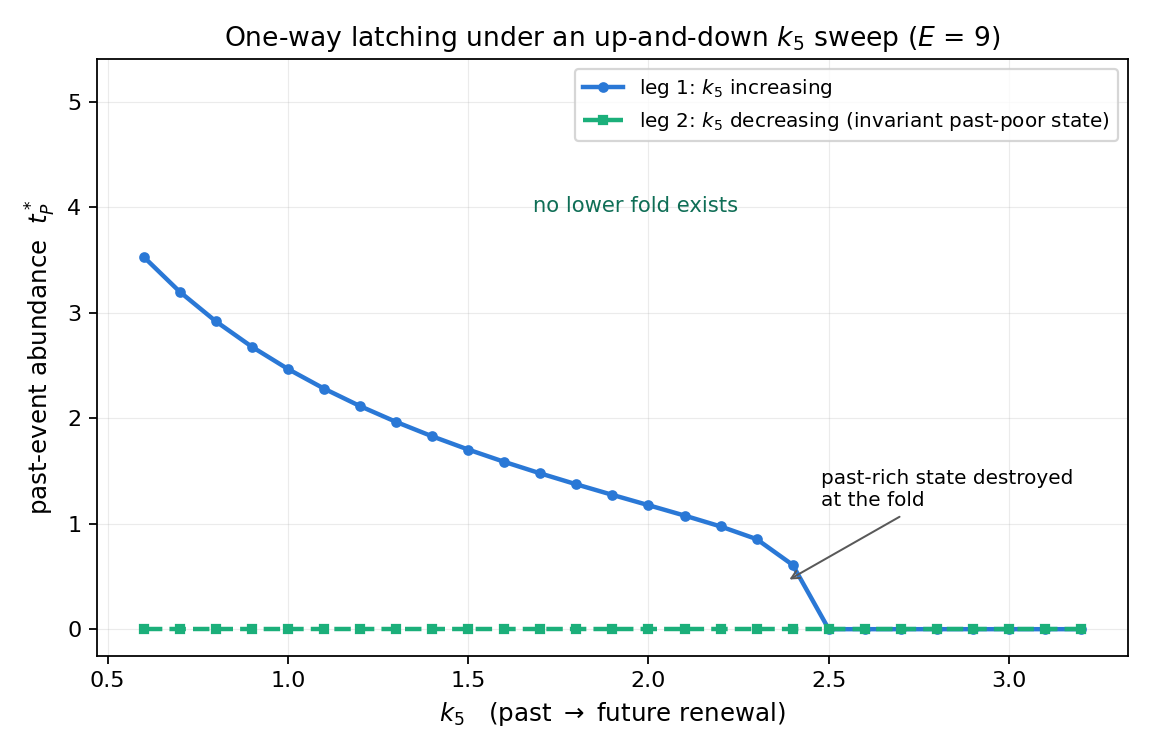}
\caption{One-way latching at $E=9$. Leg~1 (blue): $k_5$ increasing, past-rich state destroyed at the fold. Leg~2 (green, dashed): $k_5$ decreasing, on the invariant past-poor state---not the continuation of a branch that later loses stability. The legs do not close into a loop.}
\label{fig:latch}
\end{figure}

The budget cannot be swept, being fixed by initial conditions, so a parameter history must act on a rate constant; $k_5$ is the natural choice. This is an external intervention and should be labelled as such: varying a rate constant corresponds physically to varying temperature, catalyst, or some other environmental control, so the system is not autonomous during the sweep. The precise statement is that the event inventory and the modelled budget remain closed while an externally controlled kinetic parameter is varied. Figure~\ref{fig:latch} shows a quasi-static sweep at $E=9$: the past-rich state persists until $k_5\approx2.4$, where it is destroyed at the fold, and the system remains past-poor on the return leg. There is no second switching threshold, because the past-poor state is invariant and stable for every $k_5$; recovery requires reseeding with finite $\tP$. Appendix~B shows this to be the singular limit of ordinary hysteresis: restoring weak reversibility closes the loop, with the lower switching threshold moving to $k_5=0$ as the reverse rates vanish.

\subsection{Cooperative nonlinearity produces history-dependence; the irreversible limit makes it one-way}

Two distinct properties are at work here and are easily conflated, so we separate them.

The first is \emph{history-dependence}: the state realised at a given parameter value depends on how the system arrived there. The second is \emph{irrecoverability}: no reversal of the control parameter restores a stable steady state once lost. In the quasi-static steady-state sense used throughout, history-dependence requires coexistence of stable steady states over a common parameter interval; the second property is stronger. We note that path dependence can arise in other ways---through transients, delayed response, hidden variables or explicitly non-autonomous forcing---which are outside the quasi-static setting considered here.

Cooperative participation is what supplies the first. The two models contain the same irreversible elementary steps R3, R4 and R5---the same imposed local ordering, taken as a premise in \S3---yet only the cooperative model is bistable. The first-order model has no bistable interval, hence no hysteretic memory: under adiabatic tracking with nonzero perturbations, reversing the control returns the system along the same branch structure it ascended. Imposed local irreversibility therefore does not by itself generate memory.

Local irreversibility is not necessary for memory either. Appendix~B shows that the weakly reversible cooperative network---in which no step is strictly one-way---retains bistability and exhibits a conventional two-threshold hysteresis loop, which is history-dependence in the ordinary sense. What the irreversible limit does is different: as the reverse fluxes vanish the lower recovery threshold is driven to the boundary of parameter space, and the closed loop opens into strict one-way latching.

The division of labour is therefore clean. Cooperative nonlinearity creates the fold, the coexistence of attractors, and hence history-dependence. The irreversible limit does not create that history; it makes it unrecoverable by parameter reversal alone. Neither property follows from the other, and only the second is related to the premise the model imposes.

\subsection{Where the two-state structure exists}

\begin{figure}[t]
\centering
\includegraphics[width=0.72\textwidth]{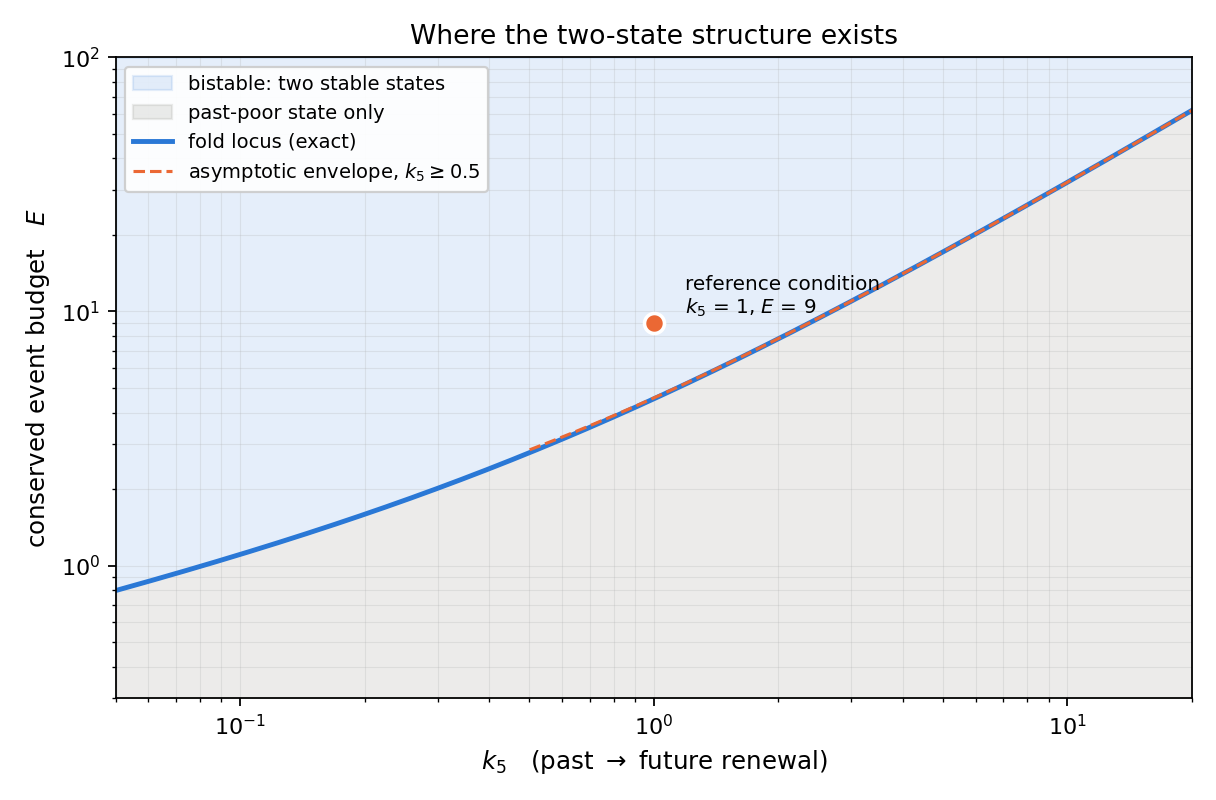}
\caption{Fold locus in the $(E,k_5)$ plane, logarithmic axes. Shaded blue: two stable steady states available. Shaded grey: past-poor state only. Dashed: the asymptotic form \eqref{eq:asym}, shown only where its relative error is below 3\%.}
\label{fig:locus}
\end{figure}

Minimising \eqref{eq:Ebranch} over $\tP$ for each $k_5$ traces the fold locus (Fig.~\ref{fig:locus}, Table~\ref{tab:locus}).

\begin{table}[H]
\centering\small
\begin{tabular}{rrrrr}
\toprule
$k_5$ & $E_{\mathrm{fold}}$ (exact) & $t_{\mathrm{P,fold}}$ & $E_{\mathrm{fold}}$ from \eqref{eq:asym} & rel.\ error\\
\midrule
0.05 & 0.7969 & 0.2815 & 1.0736 & $+34.730\%$\\
0.10 & 1.1059 & 0.3593 & 1.3104 & $+18.497\%$\\
0.20 & 1.5913 & 0.4450 & 1.7282 & $+8.601\%$\\
0.35 & 2.2108 & 0.5135 & 2.3011 & $+4.086\%$\\
0.50 & 2.7785 & 0.5536 & 2.8443 & $+2.370\%$\\
0.75 & 3.6718 & 0.5938 & 3.7151 & $+1.180\%$\\
1.00 & 4.5295 & 0.6181 & 4.5602 & $+0.678\%$\\
1.50 & 6.1887 & 0.6462 & 6.2059 & $+0.277\%$\\
2.00 & 7.8051 & 0.6619 & 7.8152 & $+0.130\%$\\
3.00 & 10.9678 & 0.6788 & 10.9710 & $+0.029\%$\\
4.00 & 14.0755 & 0.6876 & 14.0754 & $-0.001\%$\\
6.00 & 20.1982 & 0.6964 & 20.1952 & $-0.015\%$\\
8.00 & 26.2465 & 0.7008 & 26.2423 & $-0.016\%$\\
12.00 & 38.2157 & 0.7049 & 38.2106 & $-0.013\%$\\
16.00 & 50.0815 & 0.7068 & 50.0763 & $-0.010\%$\\
24.00 & 73.6346 & 0.7084 & 73.6295 & $-0.007\%$\\
32.00 & 97.0423 & 0.7091 & 97.0375 & $-0.005\%$\\
64.00 & 189.9643 & 0.7096 & 189.9603 & $-0.002\%$\\
\bottomrule
\end{tabular}
\caption{Fold locus of the cooperative model, $k_1=k_2=k_3=k_4=1$, with the asymptotic form \eqref{eq:asym}, $c_0=0.701898$, and its relative error. The expansion is a large-$k_5$ result and is not valid at the small-$k_5$ end.}
\label{tab:locus}
\end{table}

The large-$k_5$ behaviour follows analytically, with all coefficients derived rather than fitted. Retaining the terms of \eqref{eq:Ebranch} proportional to $k_5$ and minimising $k_5(1/\tP + 2\tP)$ gives $t_{\mathrm{P,fold}}\to1/\sqrt2$ and leading term $2\sqrt2\,k_5$; evaluating the remaining terms there,
\begin{equation}
E_{\mathrm{fold}} \simeq 2\sqrt2\,k_5 + \sqrt{\tfrac{3}{2\sqrt2}}\,\sqrt{k_5} + c_0,\qquad c_0 = \tfrac{1}{\sqrt2} - \tfrac{g'^2}{2f''} \approx 0.70190,\label{eq:asym}
\end{equation}
The constant term requires care. Writing $E = k_5 f(\tP) + \sqrt{k_5}\,g(\tP) + \tP$ with $f = 1/\tP + 2\tP$ and $g = [(1+\tP^2)/(2\tP)]^{1/2}$, the minimiser is displaced from $1/\sqrt2$ by $\delta \simeq -g'/(\sqrt{k_5}\,f'')$, and although $\delta$ is $O(k_5^{-1/2})$ its contribution to $E$ is $-g'^2/(2f'')$, which is of order one. With $f''=4\sqrt2$ and $g'=-1/(4g)$ at $\tP=1/\sqrt2$, this gives $c_0 = 1/\sqrt2 - g'^2/(2f'') = 0.707107 - 0.005208 = 0.701898$.\footnote{An earlier draft gave the constant as $1/\sqrt2\approx0.707107$, omitting the displacement of the minimiser. The correction was pointed out in an external critique and has been confirmed independently; the leading two terms are unaffected.} Numerically, $E_{\mathrm{fold}} - 2\sqrt2\,k_5 - 1.030\sqrt{k_5}$ tends to 0.7019 as $k_5$ grows (0.7032 at $k_5=10^3$, 0.7023 at $10^4$). The expansion is accurate to better than 0.7\% for $k_5\ge1$ and to 0.02\% for $k_5\ge3$, but reaches 35\% at $k_5=0.05$. In the large-$k_5$ regime the requirement becomes $E/k_5\gtrsim2.83$.

\subsection{Interior bistability with irreversible activation and completion}

The boundary location of the past-poor state raises a question that bears directly on the interpretation: can two stable steady states, both interior, coexist while the commitment of an event remains strictly one-way? Appendix~B obtains interior states by reversing every irreversible step, including completion, which conflicts with the postulate that an activated event, once formed, proceeds to the past. We therefore consider the minimal alteration---a partially reversible network---that keeps activation and completion strictly forward and makes only renewal reversible:
\[
2\tF \underset{k_2}{\overset{k_1}{\rightleftharpoons}} \tN,\qquad \tN + 2\tP \xrightarrow{k_3} \tS,\qquad \tS \xrightarrow{k_4} 4\tP,\qquad \tP \underset{k_{-5}}{\overset{k_5}{\rightleftharpoons}} \tF.
\]
The only new step is $\tF\to\tP$ at rate $k_{-5}\tF$. It does not undo any completion; it is the converse direction of the relational exchange of \S4.1, an event that is future for one reference being past for another, and it is a phenomenological term in the same sense as R5. Because it carries the stoichiometry of R5 reversed, the budget $E=\tF+2\tN+\tP+4\tS$ is conserved exactly, and $\tP=0$ is no longer invariant.

\begin{figure}[t]
\centering
\includegraphics[width=0.78\textwidth]{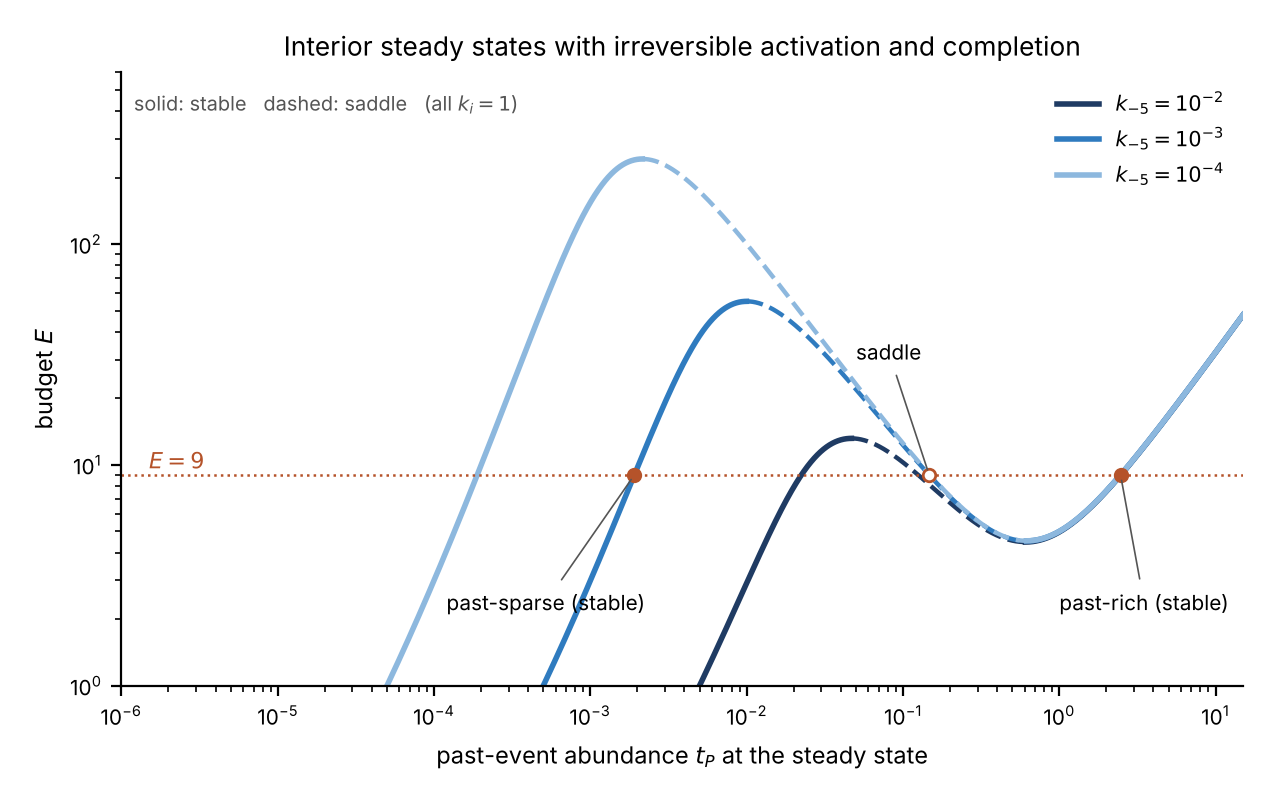}
\caption{The steady-state curve $E(\tP)$ of the network of \S6.7, computed from the exact parameterisation \eqref{eq:partialF}, for three reverse-renewal rates. Solid branches are stable and dashed branches are saddles, as confirmed by the eigenvalues at the marked states (Table~\ref{tab:partial}). The lower fold converges to the irreversible value 4.5295; the upper fold rises as $\tfrac12k_{-5}^{-2/3}$ \eqref{eq:Eup}.}
\label{fig:partial}
\end{figure}

\paragraph{Exact parameterisation.} At an interior steady state $k_3\tN\tP^2 = k_4\tS$, $k_1\tF^2 = (k_2+k_3\tP^2)\tN$ and $k_5\tP - k_{-5}\tF = 2k_4\tS$. Eliminating $\tN$ and $\tS$ gives a quadratic in $\tF$ whose unique positive root is
\begin{equation}
\tF(\tP) = \frac{2k_5\tP}{k_{-5} + \sqrt{k_{-5}^2 + 4C(\tP)\,k_5\tP}},\qquad C(\tP) = \frac{2k_1k_3\tP^2}{k_2+k_3\tP^2},\label{eq:partialF}
\end{equation}
with $\tN = k_1\tF^2/(k_2+k_3\tP^2)$ and $\tS = k_3\tN\tP^2/k_4$, all strictly positive. Every $\tP>0$ therefore determines exactly one interior steady state, and the interior steady states on a leaf are exactly the solutions of $E(\tP) \equiv \tF+2\tN+\tP+4\tS = E_0$. This reduction is exact; the number of solutions is read from the shape of the single curve $E(\tP)$ (Fig.~\ref{fig:partial}).

\paragraph{Two folds in the budget.} At unit constants $E(\tP)$ rises from zero, reaches a maximum, falls to a minimum and rises again. Its two turning points are folds. The lower fold lies at $E_{\mathrm{low}} = 4.4713$, $4.5237$, $4.5289$ for $k_{-5}=10^{-2},10^{-3},10^{-4}$ and converges to the irreversible value 4.5295. The upper fold is new. For small $\tP$ the branch reduces to $E\simeq2\tF^2$ with $\tF\simeq2\tP/(k_{-5}+\sqrt{k_{-5}^2+8\tP^3})$; writing $\tP=k_{-5}^{2/3}u$, this is $E\simeq k_{-5}^{-2/3}\,g(u)$ with $g(u)=8u^2/(1+\sqrt{1+8u^3})^2$, whose maximum is $g(1)=\tfrac12$. Hence
\begin{equation}
E_{\mathrm{up}} \simeq \tfrac12\,k_{-5}^{-2/3},\qquad t_{\mathrm{P,up}} \simeq k_{-5}^{2/3}\qquad (k_i=1),\label{eq:Eup}
\end{equation}
which the exact branch confirms: $E_{\mathrm{up}} = 13.18$, $55.02$, $242.86$, $1100.4$ at $k_{-5}=10^{-2},10^{-3},10^{-4},10^{-5}$, against $10.77$, $50.00$, $232.08$, $1077.2$ from \eqref{eq:Eup}, the ratio approaching one with a relative correction of order $k_{-5}^{1/3}$. For $E_{\mathrm{low}}<E<E_{\mathrm{up}}$ there are three interior steady states; outside that window there is one. The stable--saddle--stable classification is demonstrated by eigenvalues at the tabulated parameter values; along the branches it is inferred from the sign of $\dd E/\dd\tP$, which matches the eigenvalues wherever they were evaluated, and is not proved for every parameter set.

\begin{table}[H]
\centering\small
\begin{tabular}{lccccc}
\toprule
 & $\tP$ & $\tF$ & $\tN$ & $\tS$ & leaf eigenvalues\\
\midrule
past-sparse (stable) & 0.001912 & 1.88577 & 3.55613 & $1.30\times10^{-5}$ & $-8.544,\ -1.247,\ -0.780$\\
saddle & 0.147139 & 1.85150 & 3.35540 & 0.07264 & $-8.416,\ -4.205,\ +0.217$\\
past-rich (stable) & 2.466378 & 1.19801 & 0.20263 & 1.23259 & $-10.714,\ -2.581\pm1.237\,i$\\
\bottomrule
\end{tabular}
\caption{Interior steady states of the network of \S6.7 at $E=9$, $k_{-5}=10^{-3}$, all other $k_i=1$. Residuals of all four balances are below $10^{-15}$. The completion flux $k_4\tS$ is positive in both stable states.}
\label{tab:partial}
\end{table}

\paragraph{What this establishes.} At $E=9$ and $k_{-5}=10^{-3}$ the three interior states are listed in Table~\ref{tab:partial}: two stable states separated by a saddle steady state, the saddle having a single positive eigenvalue. They differ little from the irreversible values of \S6.2 in the past-rich state ($\tP=2.4664$ against 2.4654) and in the saddle (0.1471 against 0.1492), because activation and completion are unchanged; the boundary state of \S6.2 becomes a past-sparse interior state. Swept in $k_5$ at $E=9$, the network shows two folds: the upper near $k_5=2.38$, as in the irreversible network, and a lower one at $k_5\simeq(8k_3\tN\tF k_{-5})^{1/2}\approx7.3\,k_{-5}^{1/2}$, with $\tF,\tN$ at the past-poor state. This is a leading-order asymptotic result as $k_{-5}\to0$, not an exact threshold at finite reverse rate; the thresholds located numerically are 0.727, 0.232 and 0.074 at $k_{-5}=10^{-2},10^{-3},10^{-4}$, i.e.\ $7.27$, $7.33$ and $7.39$ times $k_{-5}^{1/2}$. The leading-order derivation is that of Appendix~B with $\epsilon\kappa_{-5}$ replaced by $k_{-5}$, since the reverse renewal flux is the only reverse term that enters it at leading order. The budget folds add a feature absent from the irreversible network: above $E_{\mathrm{up}}$ the past-sparse state no longer exists, so a sufficiently large budget forces ignition. As $k_{-5}\to0$, $E_{\mathrm{up}}\to\infty$ and the irreversible result, a past-poor state stable at every budget, is recovered.

Two levels of irreversibility should be kept apart. At the level of the event, nothing is reversed: an activated event has no route back to the now class, its only transition is completion, and in both stable states activation and completion balance at a positive flux, $k_3\tN\tP^2 = k_4\tS>0$. At the level of the network, the two stable states are two persistent compositions with different completion fluxes, $1.3\times10^{-5}$ and $1.23$ here; they are not two directions of progression. The network of this subsection is therefore the physically admissible realisation of the bistability result: both steady states are interior, as the relational reading requires, and the commitment of every event to the past remains strictly one-way.

These results were obtained from the exact parameterisation \eqref{eq:partialF} evaluated at 30-digit precision, with stability from the Jacobian of the full system, the neutral eigenvalue normal to the leaf excluded. They were computed by the same AI system that produced the other calculations (Appendix~A).

\section{The activated complex and the fast-intermediate limit}

\subsection{The activated class as transition state}

In transition-state theory (TST) a reaction proceeds from reactants, through a pre-equilibrium, to a transient high-energy configuration---the transition state, or activated complex---at the maximum of energy along the reaction path, and thence to products~\cite{eyring1935,evanspolanyi1935}. The energy difference between reactants and activated complex is the activation energy; configurations that never acquire it never reach the complex. A complex that has formed may still revert to reactants, the fraction that proceeds being the transmission coefficient; but a system carried past the dividing configuration proceeds to products. Transition-state configurations of this kind have been observed spectroscopically, for example in the catalytic oxidation of CO on a ruthenium surface by femtosecond X-ray absorption~\cite{ostrom2015}.

Under the energy reading of \S4.4 the cooperative network has this architecture, and it is the activated class $\tS$, not the now class $\tN$, that occupies the transition-state position:
\begin{itemize}
\item $\tN$ is an energy state of intermediate potential energy (two units), formed by converting the kinetic energy of two future events, and in reversible exchange with future events through R1/R2. It is the reactant side of a pre-equilibrium, not a point of commitment.
\item $\tS$ is the activated energy state: the highest energy per event (four units), the peak of potential energy, reached only when a now event gathers the activation energy $E_a=2$ units as kinetic energy from two past events, and leaving the class by completion R4 into four past events. It is the point of commitment to the past.
\item $\tP$ is the product: dispersed, completed, and---in the strictly irreversible model---unable to re-form the complex from which it came.
\end{itemize}
On the relational reading of \S2, the present is the moment of commitment, instantaneous and irreversible; the programme accordingly identifies the activated class with the present. That identification is part of the interpretive postulate of category~D and is used nowhere in the analysis.

The correspondence extends to reversion. In the strictly irreversible model R3 has no reverse, so every complex that forms completes: the forward-commitment probability is one. In the weakly reversible network of Appendix~B an activated event can either revert ($\tS\to\tN+2\tP$, at rate $\epsilon\kappa_{-3}$) or complete ($\tS\to4\tP$, at rate $k_4$), so the probability that a formed complex proceeds to products is
\begin{equation}
\pi_{\to} = \frac{k_4}{k_4 + \epsilon\,\kappa_{-3}},\label{eq:transmission}
\end{equation}
the analogue of the transmission coefficient. At $\epsilon=10^{-3}$ and unit constants $\pi_{\to}=0.999$; as $\epsilon\to0$ it tends to one, recovering the irreversible model as the limit in which every complex that forms goes on to products.

What the network does not supply should be stated with equal precision. The activation energies \eqref{eq:Ea}--\eqref{eq:Eapp} are defined under the postulated division of energy into kinetic and potential forms; the network contains no potential-energy surface over a continuous reaction coordinate, no dividing surface, no partition functions and no temperature, so the rate constants are not obtained from an Eyring expression; they remain free parameters read as ensemble propensities (\S3.3). The correspondence is structural: the network has the kinetic architecture that transition-state theory describes, with energy content per event supplying the ordering of states. At the reference parameters the activated class is a populated intermediate; only in the fast-intermediate limit below does it approach the fleeting complex of idealised transition-state theory.

\subsection{The fast-intermediate limit}

A transition state is fleeting. At the reference parameters of \S6, however, it is not: at $k_4=1$ the activated class holds 55\% of the budget, a populated intermediate rather than a transient one. The transition-state reading is therefore realised exactly in the singular fast-intermediate limit $k_4\to\infty$~\cite{tikhonov1952,segel1989}, in which the activated-now variable vanishes while its flux remains finite: $\tS = k_3\tN\tP^2/k_4\to0$ with $k_4\tS$ finite. In this limit, and with R1/R2 in the stiff quasi-equilibrium of \S5.5, the cycle turnover takes the transition-state form
\begin{equation}
J = k_3\,K_1\,\tF^2\,\tP^2,\qquad K_1 = k_1/k_2,
\end{equation}
a rate constant composed with a pre-equilibrium constant, multiplying the reactant abundances.

Eliminating $\tS$ merges R3$'$ and R4$'$ into $\tN + 2\tP\to4\tP$, reducing the system to three variables with budget $E=\tF+2\tN+\tP$. The steady-state condition \eqref{eq:factor2} is unchanged, so the fold, the two stable steady states, the ignition threshold and the latching all survive.

\subsection{A reciprocal symmetry, and its condition}

For general constants the reduced branch equation is
\begin{equation}
E(\tP) = \sqrt{\frac{k_5\,(k_2 + k_3\tP^2)}{2k_1k_3\tP}} + \frac{k_5}{k_3\tP} + \tP.\label{eq:reduced}
\end{equation}
Consider the map $\tP\to\alpha/\tP$ with parameter $\alpha$; its positive fixed point is $\tP=\sqrt\alpha$. Requiring $E(\alpha/\tP) = E(\tP)$ forces $\alpha = k_5/k_3$ from the second and third terms and $\alpha = k_2/k_3$ from the radical. Both hold simultaneously if and only if
\begin{equation}
k_2 = k_5,\label{eq:k2k5}
\end{equation}
whereupon \eqref{eq:reduced} is exactly invariant under $\tP\to(k_5/k_3)/\tP$ and the fold sits at the self-reciprocal point $t_{\mathrm{P,fold}} = (k_5/k_3)^{1/2}$. We verified this numerically: at $(k_1,k_2,k_3,k_5) = (2.3,0.7,1.9,0.7)$ the symmetry holds to machine precision with $t_{\mathrm{P,fold}} = 0.606977 = \sqrt{0.7/1.9}$, and at $(0.4,2.5,3.1,2.5)$ with $t_{\mathrm{P,fold}} = 0.898026$; at $(1,1,1,2)$, where $k_2\neq k_5$, it fails by 0.61 in $E$.

\begin{figure}[t]
\centering
\includegraphics[width=0.72\textwidth]{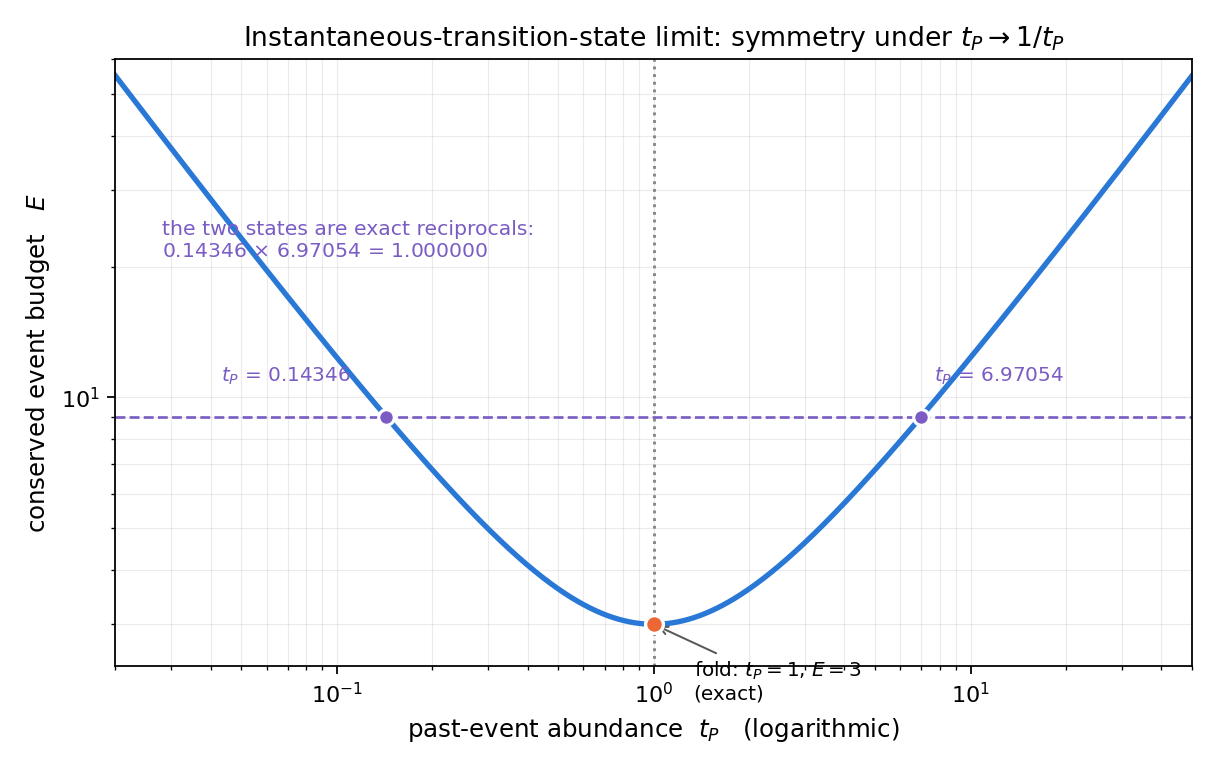}
\caption{Reduced branch locus at unit parameters, plotted against $\log\tP$. The curve is exactly symmetric about $\tP=1$, placing the fold at $\tP=1$, $E=3$ and making the two steady-state values at any budget above the fold exact reciprocals. Under condition \eqref{eq:k2k5} with general constants the symmetry centre, the fixed point $\sqrt\alpha$ of the map with $\alpha=k_5/k_3$, moves to $\sqrt{k_5/k_3}$.}
\label{fig:recip}
\end{figure}

The symmetry is therefore a property of the reduced non-dimensional branch equation under a specific parameter relation, not a generic feature. Condition \eqref{eq:k2k5} has a natural reading: $k_2$ and $k_5$ are the two routes by which event activity returns to future status---directly from now events, and by renewal from past events---so the symmetry holds exactly when those two routes are equally fast. For unit parameters, $t_{\mathrm{P,fold}}=1$ and $E_{\mathrm{fold}}=3$ exactly, and the saddle and past-rich states at any budget are exact reciprocals: at $E=9$ they are 0.14346 and 6.97054, whose product is 1.000000 (Fig.~\ref{fig:recip}). We state this as a structural property of the branch equation and draw no physical conclusion from it.

One feature of the symmetry deserves note, since it bears on \S6.5. The reciprocal partners are not two attractors: at $E=9$ they are the saddle at 0.14346 and the stable steady state at 6.97054, lying respectively below and above the self-reciprocal fold point. The map $\tP\to(k_5/k_3)/\tP$ therefore exchanges attractor with separatrix. It is a symmetry of the stationary locus and not of the flow upon it: the set of stationary abundances is reciprocally symmetric, while dynamical stability breaks that symmetry. Both members of a pair are equilibria and either can in principle be occupied; what differs is stability and generic accessibility, since trajectories approach the attractor from an open set of initial conditions whereas reaching the saddle requires preparation on its stable manifold.

\subsection{Planar dynamics}

With three variables and one conservation law each invariant leaf is two-dimensional. Strange chaotic attractors are therefore excluded, and bounded $\omega$-limit sets are restricted by the Poincar\'e--Bendixson theorem to equilibria, periodic orbits, or sets composed of equilibria together with connecting trajectories~\cite{perko2001}. This restriction is unavailable in the four-variable system.

One feature is lost in the reduction: at $E=9$ the eigenvalues become $-8.544$ and $-47.59$, both real, in place of the complex pair $-2.580\pm1.236\,i$. The damped oscillation of \S6.2 was supplied by the intermediate, so if the ringing is regarded as meaningful the activated variable must retain finite lifetime and the reduction is not available. The choice between a fleeting activated complex, as transition-state theory idealises it, and a populated one with finite lifetime is therefore not merely interpretive: it decides whether the approach to the running state rings.

\section{Relation to the published model}

The published model's dynamical content is not thereby lost. What it exhibits, once the yield is corrected, is a transcritical ignition threshold: below a critical event budget the progression cannot sustain itself, above it the past-rich state grows from any perturbation whatever. Note that in exact deterministic arithmetic $\tP=0$ remains invariant even after it becomes linearly unstable, so statements about the system leaving the past-poor branch presuppose a nonzero perturbation; the branch geometry, not the seeding, is what distinguishes the two models. That is a genuine threshold phenomenon in a conserved relational network. What it does not exhibit is two coexisting stable steady states with history dependence, and obtaining those requires cooperative participation.

\begin{table}[H]
\centering\small
\begin{tabular}{>{\raggedright\arraybackslash}p{3.1cm}>{\raggedright\arraybackslash}p{4.3cm}>{\raggedright\arraybackslash}p{6.0cm}}
\toprule
Feature & Published model~\cite{boman2025} & Present analysis\\
\midrule
Activated-state yield & $\tS\to2\tP$ & $\tS\to3\tP$ required by closure for the first-order model; $\tS\to4\tP$ for the cooperative one\\
Bistability criterion & simultaneous partial balances & two complete stable equilibria plus an intervening saddle, on a fixed leaf\\
Autocatalytic order & first order in $\tP$ & first order tested and shown insufficient for every yield $n$\\
Bistable mechanism & not established & second order in $\tP$\\
Method & algebraic scenarios & Jacobians, bifurcation geometry, Radau integration\\
Activated state & finite populated intermediate & also examined in the fast-intermediate limit, where it is the activated complex of transition-state theory\\
Meaning of the budget & not specified & exactly conserved weighted count; conserved energy under the stated postulate (\S4.4)\\
\bottomrule
\end{tabular}
\caption{What changes between the published analysis and the present one.}
\end{table}

\section{Discussion}

\paragraph{Closure selects the stoichiometry.} Requiring a positive linear conserved event-weight functional is not decorative: given the scheme and the normalisation $w_F=1$ it forces $n=m+2$, and it fixes the stoichiometric multipliers in the balances. The distinction between reaction order and stoichiometric coefficient is material---order governs the form of the flux, the multiplier its weight in each balance---and conflating them yields a system that manufactures budget at a rate proportional to a flux. The energy reading of \S4.4 sharpens this rather than relaxing it: if the weights are energy per event, the multipliers state how much energy each transition moves, and altering one would create or destroy energy.

\paragraph{Conservation is not the obstacle; first-order participation is.} A closed, exactly conserved network can fold, latch and exhibit finite-amplitude ignition. What it cannot do, within this network class, is fold while the autocatalytic step is first order in $\tP$, and that holds for every decomposition yield. The kinetic content is \eqref{eq:Jlin}: with $\tN$ pinned, turnover is linear in $\tP$. The single structural change that removes the obstruction is cooperative activation.

\paragraph{The fold requires nonequilibrium kinetic asymmetry, not strict irreversibility.} Three regimes should be distinguished, and only the third excludes the fold.

Under strictly irreversible kinetics the fold is present and the latch is one-way (\S6.2, \S6.4). Under weakly reversible but not detailed-balanced kinetics the fold survives, with the bistable interval intact and the latch closing into a two-threshold hysteresis loop (Appendix~B); no step is one-way here, so strict irreversibility is plainly not what the fold requires. Under detailed-balanced kinetics the fold disappears: the network admits a single Wegscheider condition, $K_1K_3K_4K_5^2 = 1$, and imposing it leaves a unique positive equilibrium in each positive compatibility class, attracting for positive initial conditions, since a detailed-balanced mass-action system possesses a strictly convex free-energy Lyapunov function~\cite{hornjackson1972,feinberg1987}.

What the fold requires is therefore nonequilibrium kinetic asymmetry---a nonzero cycle affinity in the chemically interpreted analogue---of which strict irreversibility is only the extreme case. The bistability reported here is not equilibrium bistability, and the asymmetry that supports it is a premise of the model rather than a consequence of it.

\paragraph{Entropy: what is established, and what is proposed.} The fold requires a nonzero cycle affinity, and a closed network must say how that asymmetry is sustained without an external reservoir. Three things should be kept distinct: stationarity of the coarse-grained composition, entropy production, and entropy exchange.

What is established is exact but limited. Around the single cycle of Appendix~C---R1, R3$'$, R4$'$ and twice R5---both the budget and the number of events are conserved: condensation changes the event count by $-1$, activation by $-2$, completion by $+3$, renewal by $0$, and the cycle vector $(1,1,1,2)$ sums these to zero. At a running steady state the event-class abundances, and any function defined solely by them, are stationary, while step by step energy is concentrated into fewer events (condensation, activation) and dispersed into more (completion). This does not establish vanishing thermodynamic entropy production. On the standard accounting of nonequilibrium thermodynamics, a mass-action steady state with nonzero cycle flux and affinity has positive entropy production, balanced in an open system by export to an environment~\cite{schnakenberg1976}. The model does not define an entropy functional, a temperature, or a multi-reference exchange mechanism, so it cannot decide the question.

What is proposed is the programme's hypothesis, stated as such. Entropy increases locally at each completion, at every reference, and decreases locally where energy is gathered into an activated event; globally, the relational exchange of \S4.1---the past of one reference being the future of another---plays the role that an environment plays for an open system, so that the change of entropy of the whole would be zero and the second law, $\Delta S\ge0$, would be met with equality. If the hypothesis held, no low-entropy initial condition would be required, in keeping with the view that the arrow of time is perspectival rather than an absolute boundary condition~\cite{rovelli2017}. Paper~1 draws no thermodynamic conclusion here, and nothing in the present network demonstrates the hypothesis: it requires an entropy functional and the explicit multi-reference model deferred below, and it departs from standard thermodynamics, in which internal exchanges between subsystems cannot cancel positive irreversible entropy production in an isolated system. The hypothesis therefore requires either that the exchange between references not be an internal exchange in that sense, or a revision of the standard accounting; which, if either, holds is not decided here. It is offered as a hypothesis for a thermodynamically consistent multi-reference model, with an explicit entropy functional, to test.

\paragraph{What creates memory, and what makes it one-way.} Section~6.5 separates two properties usually run together. Cooperative second-order participation creates the fold and the coexistence of attractors, and hence history-dependence; the two models share the same irreversible elementary steps, so imposed local ordering plainly does not generate memory by itself. Nor is local irreversibility necessary for memory: the weakly reversible network of Appendix~B is bistable and shows an ordinary two-threshold hysteresis loop. The role of the irreversible limit is to drive the lower recovery threshold to the boundary $k_5=0$, converting recoverable hysteresis into one-way latching. The threshold scales as the square root of the reverse rate: $k_{5,\mathrm{lower}}\simeq(8k_3\tN\tF k_{-5})^{1/2}$ in the partially reversible network of \S6.7 and $k_{5,\mathrm{lower}}\simeq(8k_3\tN\tF\epsilon\kappa_{-5})^{1/2}$ in the fully reversible network of Appendix~B, with $\tF$, $\tN$ at the past-poor state in each case; the degree of recoverability accordingly varies continuously with the reverse-rate scale in this family of models. Note also that the budget is exactly conserved throughout, so the one-wayness involves no depletion: it is a property of basins and their boundaries, not of a resource running down.

\paragraph{Seeding is not a well-formed notion here.} The direction of the irreversible transition should be stated plainly. What the cooperative model makes irrecoverable is the loss of the past-rich state: raising $k_5$ past the fold extinguishes the progression, and lowering it again does not restore it. In the other direction the past-poor state is stable at every budget, so on a naive reading ignition would require a finite seed of past events introduced into a network containing none.

That reading is not available under the ontology of \S2. If the labels future, now and past are relational---if no event possesses its status in itself, but only with respect to the others---then a configuration in which one class is empty is not a state of the system but a degeneracy of the labelling scheme. There is nothing for the remaining events to be future or now relative to. A ``seed'' would be a transition from that degenerate configuration into a populated one, and since the configuration is not a physical state of the network, the transition is not a physical process that the model owes an account of. The physical domain is accordingly the interior, in which every class is populated.

The dynamics corroborate this rather than merely tolerating it. The set $\tP=\tS=0$ is invariant, and sub-threshold trajectories approach it exponentially at rate $k_5$ without attaining it: from $\tP(0)=0.30$ at $E=9$ the past-event abundance falls to $3\times10^{-9}$ by $s=30$ and $2\times10^{-17}$ by $s=50$, and is nonzero at every finite $s$. No interior trajectory reaches the boundary, and none departs from it. Ontology and dynamics therefore agree: the network never leaves the domain in which its labels are meaningful, and never needs to enter it.

Two consequences follow for how the results should be read. The coexistence established in \S6.2 is between an interior attractor and a boundary attractor, and only the first is a physical state on the present reading; what the second represents is unbounded decay of the progression toward the degenerate limit, not an alternative regime the network can occupy. And the question of origination does not arise, since the analysis concerns a network whose classes are all populated---which, for the intended reading, is the case of interest. Given such a network, cooperative participation is what sustains progression against perturbation, the fold marks the budget below which it cannot be sustained, and crossing that fold sends the system into irreversible decay.

\paragraph{Strict one-wayness is an idealisation, and the ontology says so.} One consequence of the reading just given deserves stating, since it qualifies the headline result. In the strictly irreversible model the past-poor attractor lies on the boundary of the positive orthant---the configuration that \S2 declares a degeneracy of the labelling rather than a state. If that reading is taken seriously, the strictly irreversible model is the one whose second attractor is not a physical state, and the network of \S6.7 is the better-behaved variant: with only renewal made reversible, every steady state is strictly interior, with all four classes populated, while activation and completion remain one-way. At $E=9$ and $k_{-5}=10^{-3}$ the three states carry $\tP = 1.9\times10^{-3}$, $1.5\times10^{-1}$ and $2.5$, none of them on the boundary. The fully weakly reversible network of Appendix~B gives interior states too, but only by also reversing completion.

It is worth naming the two readings separately. \emph{Mathematical bistability} is the coexistence, in the strictly irreversible system, of an interior attractor with a stable boundary equilibrium. \emph{Physically admissible bistability}, under the relational requirement that all classes remain populated, is the coexistence of two interior stable steady states; \S6.7 realises it with activation and completion strictly one-way. The strictly irreversible network establishes the bifurcation mechanism in its sharpest mathematical form; the network of \S6.7 supplies the physically admissible realisation, and Appendix~B a fully reversible comparison. The qualitative fold-and-separatrix structure derived in \S6 persists in the partially reversible network of \S6.7 and converges to the strictly irreversible geometry as $k_{-5}\to0$, but it is only the qualitative structure that carries over: the equilibrium locations, the separatrix, the ignition threshold and the lower switching threshold all vary with $k_{-5}$, and the past-poor boundary equilibrium becomes a past-sparse interior one. In the network of \S6.7 at $k_{-5}=10^{-3}$ and $E=9$, the saddle sits at $\tP=0.147139$ against 0.1492 in the irreversible limit, the past-rich state at 2.466378 against 2.4654, and the lower fold budget at 4.523717 against 4.5295. (The fully reversible network of Appendix~B gives different values at $\epsilon=10^{-3}$: 0.1473, 2.4205 and 4.5263.)

The cost is that latching is then not strictly one-way, even though every event's commitment is. In both the network of \S6.7 and that of Appendix~B a recovery threshold exists at $k_{5,\mathrm{lower}}\propto(\text{reverse rate})^{1/2}$---$(8k_3\tN\tF\epsilon\kappa_{-5})^{1/2}$ in Appendix~B, $(8k_3\tN\tF k_{-5})^{1/2}$, asymptotically $\approx7.3\,k_{-5}^{1/2}$ at $E=9$, in \S6.7--- so a route back from decay is available in principle, though it requires driving the renewal rate below a value that shrinks only as $\sqrt\epsilon$. The honest formulation is that one-wayness in this network is scale-dependent rather than absolute: strict irreversibility is the $\epsilon\to0$ idealisation, and what survives at any finite reversibility is a recovery threshold pushed arbitrarily close to the boundary of parameter space without reaching it. Whether a physical realisation should be modelled at $\epsilon=0$ or at small $\epsilon>0$ is not settled here, and the two differ in interpretation rather than in any result reported above.

\paragraph{Two levels of barrier.} The results admit a landscape picture, provided two levels are kept apart. At the level of the ensemble, the two stable steady states are two basins and the saddle of \S6.2 is the ridge between them; crossing it is what ignition and extinction require. Because the system is not detailed-balanced it possesses no true potential function, so the landscape is a quasi-potential in the sense of large-deviation theory~\cite{freidlin2012}---a representation of relative stability, not of energy. At the level of the individual event, each transition passes through the activated complex $\tS$ of \S7, the transition state of a single progression step. The ensemble saddle and the event-level transition state are different objects at different scales: the first separates the stable steady states of the whole network; the second is traversed by every event in either regime.

\paragraph{Local bistability and the Universe as a whole.} The programme reads the network as running at every point of reference, each relative to all others. The present analysis treats one well-mixed ensemble---the network at a single reference---and the bistability proved is a property of that ensemble. Whether, and in what form, it persists for a population of mutually coupled references is a question about a spatially extended system, and it is not answered here. Known results for coupled bistable media indicate what to expect. Each reference retains its local bistability; globally, regions in either regime can form, separated by fronts that propagate as trigger waves~\cite{fife1977}, one regime tending to invade the other. A conserved total budget can change the outcome: as an invading front consumes the shared pool it can stall, and the two regimes can then coexist indefinitely in different regions, a mechanism known in mass-conserved bistable systems as wave-pinning~\cite{mori2008}. These are expectations from analogous systems, not results for the event network, and establishing them requires the spatial ontology that the next paragraph defers.

\paragraph{What is not claimed.} The work is entirely theoretical. The model does not derive temporal irreversibility. It assumes an irreversible local ordering of relational event transitions and asks what nonlinear dynamics follow. This is consistent with, and dependent on, the companion framework's own position: there the emergence of the arrow from asymmetries among event relations---including irreversible interactions---is explicitly left to a complete theory~\cite{boman2026rc}, and nothing established here closes that gap. What the present analysis adds is conditional: given an irreversible ordering, the minimal event-interaction order required, within this specified network, for two coexisting stable steady states is second order in $\tP$, and first order is insufficient for every decomposition yield.

Nor does it establish that time itself is bistable: the results are properties of the proposed network under the closure hypotheses of \S3. The two states differ in how event activity is distributed among relational classes, not in the rate at which anything flows. The identification of the budget with energy, of the activated class with the transition state and the present, of the rate constants with ensemble propensities, and of renewal with exchange between references are interpretive postulates, and the relational account of entropy is a hypothesis; the paper states them, uses them to read its results, and derives nothing from them.

\paragraph{Open directions.} The stiff quasi-equilibrium limit of \S5.5 permits analytic elimination of $\tN$ and a further reduction. The quantitative map from budget units to physical energy is not supplied, and supplying it is the step that would turn the energy reading from a postulate into a testable claim. A multi-reference network, in which renewal at one reference is resolved into exchange with others, together with an entropy functional, is the step that would test the entropy hypothesis given above. Spatial extension, in which event classes are allowed to propagate, likewise requires an explicit spatial ontology---what propagates, why the classes share a propagation law, and what the spatial coordinate represents relative to the change parameter $s$---and is deferred; preliminary computations indicate trigger-wave behaviour of the kind familiar from bistable reaction--diffusion systems~\cite{fife1977}, but are not reported here for that reason.

\section{Conclusion}

We have identified the structural condition separating soft ignition from bistable progression in this conserved event network. First-order participation of the past-event variable leaves it as a common factor in the steady-state condition, pins the now-event abundance on the progressing branch, and permits only a transcritical exchange of stability---and this holds for every completion yield, so it is not an artefact of the value that closure selects. Cooperative second-order participation removes the obstruction: the branch folds, a separatrix appears, ignition becomes finite-amplitude, and the realised state acquires memory. Conservation therefore does not prevent bistability. What decides it is the nonlinear order by which accumulated events influence their own continuation.

The analysis also separates two effects that are commonly conflated. Cooperative nonlinearity creates bistability and history-dependence; nonequilibrium kinetic asymmetry controls only recoverability. When only renewal is made reversible, both stable steady states lie in the interior while activation and completion stay strictly one-way, and the states form an ordinary hysteresis loop, in the renewal rate and in the budget alike. Strict irreversibility drives the lower recovery threshold to the boundary---as $k_{5,\mathrm{lower}}\simeq(8k_3\tN\tF k_{-5})^{1/2}$ in the partially reversible network of \S6.7, and as $(8k_3\tN\tF\epsilon\kappa_{-5})^{1/2}$ in the fully reversible network of Appendix~B---producing one-way latching as a singular limit. The irreversible model is thus not a separate mechanism but the limiting geometry of a reversible nonequilibrium family---and, since the relational reading requires every class to remain populated, it is that interior member of the family that supplies the physically admissible realisation.

Read through the stated postulate, the same mathematics describes progression as a sequence of energy transitions within an exactly conserved total: parcels condense, concentrate in an activated complex at the top of the energy ladder, and disperse into the past, at rates that are ensemble propensities in Boltzmann's sense, with the composition stationary at the running state while energy is concentrated and dispersed locally. That reading adds meaning, not results; the results stand without it.

None of this derives time or its arrow; the relational ordering and the irreversibility of local transitions were assumed throughout. What the analysis shows is what such an ordering requires in order to persist.

The significance is methodological as much as interpretive. A relational ontology does not by itself determine kinetics, conservation, or the direction of progression; each remains an explicit hypothesis, stated in \S3 and carried no further than the equations support. But once those hypotheses are fixed, bifurcation analysis settles exactly which proposed mechanisms can support distinct persistent stable steady states and which cannot. For the network studied here the answer is sharp: first-order participation is insufficient, cooperative participation is minimal, and strict one-wayness is an ideal limit rather than the origin of memory. That is a corrected dynamical foundation for the earlier event-progression model~\cite{boman2025}, and a design principle for closed, driven networks in which conserved activity amplifies its own continuation.

The central lesson is that conservation does not silence nonlinear history. Once participation becomes cooperative, a closed network can remember which stable steady state it has entered; irreversibility determines only how difficult that memory is to undo.

\appendix

\section{Numerical methods}

Steady states were obtained in closed form from the parameterisations of \S5.1 and \S6.2 and verified by residual, $\max|\dd y/\dd s|\le4\times10^{-16}$. Trajectory integrations use implicit Radau IIA with analytic Jacobian, rtol $=10^{-11}$, atol $=10^{-13}$; the ignition threshold was bisected to 45 iterations, giving $t^*_{\mathrm P} = 0.371404$.

Conservation of the budget is not a convergence diagnostic. For a linear invariant $\mathbf{w}^\top x$ with $\mathbf{w}^\top f\equiv0$, explicit Euler gives $\mathbf{w}^\top x_{n+1} = \mathbf{w}^\top x_n$ exactly, so the budget is preserved regardless of step size: integrating from $E(0)=9$ to $s=10$ gives $\tP=2.465297$ at $h=10^{-4}$ but $2.464744$ at $h=5\times10^{-2}$, with $E$ constant to $7\times10^{-15}$ throughout, and at $h=0.2$ the integration diverges while $E$ remains exact until it does. The property is not peculiar to Euler. Any Runge--Kutta method advances by a linear combination of flux evaluations, each annihilated by $\mathbf{w}^\top N = 0$, so all such methods preserve a linear invariant exactly; we confirmed this for RK45, DOP853, Radau, BDF and LSODA, each holding $E$ to $\le6\times10^{-15}$ over the same integration. This is standard structure preservation rather than a result of the present work, and we record it only because it disqualifies an otherwise natural check. Tracking $\Delta E$ is blind to numerical accuracy: it certifies only that the update respects $\mathbf{w}^\top N=0$, and continues to hold while the trajectory diverges---including for the adaptive solvers on which the reported trajectories rest. Step-size validation must rest instead on local error control or on systematic refinement. We used the latter: refinement against a Radau reference gave error ratios 2.003, 2.001, 2.001, 2.000, 2.000 under successive halving, consistent with first-order convergence, and the adaptive Radau runs on which all reported trajectories rest carry their own embedded local error estimate. Control coefficients used symmetric logarithmic differencing, $\varepsilon=0.01$--$0.02$, checked against the summation theorem to $\pm0.0005$. No clipping was applied; all variables remained non-negative unaided.

\paragraph{Independent re-implementation.} For version~3 the key results were recomputed with code written independently of the archived scripts and using different methods: exact symbolic algebra (SymPy) for the conservation laws, the transcritical threshold \eqref{eq:Ec}, the reciprocal symmetry and the cycle and conservation spaces of Appendix~C; Newton iteration from 3000 random starts on the leaf for the steady states; LSODA in place of Radau for the ignition threshold; and closed-form steady states with central differencing at step $10^{-5}$ for the control coefficients. All agreed with the values reported: $E_c=1.7071068$; the three $m=2$ states at $E=9$ and their leaf eigenvalues to the precision shown; $t_{\mathrm{P,fold}}=0.6181$, $E_{\mathrm{fold}}=4.5295$; $t^*_{\mathrm P}=0.3714$; control coefficients summing to $1$ to $10^{-11}$; the reduced reciprocal pair $0.14346\times6.97054=1.000000$; and the coefficient of \eqref{eq:lower}. The lower fold of Appendix~B, located independently by bisection on the number of roots, agreed with the reported values to within 0.15\% (0.23112 against 0.23079 at $\epsilon=10^{-3}$). The same check confirmed the correction to the constant of \eqref{eq:asym}. The results of \S6.7 were computed directly from the exact parameterisation \eqref{eq:partialF} and the Jacobian; their fold locations were confirmed by locating the turning points of $E(\tP)$ by bounded minimisation to $10^{-14}$ in $\tP$. Multistart searches give strong evidence that all steady states have been found but do not by themselves prove it; for the irreversible networks the closed-form parameterisations of \S5.1 and \S6.2 do.

Unless stated otherwise, $k_1=k_2=k_3=k_4=k_5=1$ and, where a budget is specified, $E=9$.

\section{Weak reversibility closes the latch into a loop}

This appendix reverses every irreversible step, including activation and completion, and serves as a comparison for the network of \S6.7, in which only renewal is reversed. Results are not transferred between the two networks without rederivation.

The reversible network is
\[
2\tF \underset{k_2}{\overset{k_1}{\rightleftharpoons}} \tN,\qquad \tN + 2\tP \underset{\epsilon}{\overset{k_3}{\rightleftharpoons}} \tS,\qquad \tS \underset{\epsilon}{\overset{k_4}{\rightleftharpoons}} 4\tP,\qquad \tP \underset{\epsilon}{\overset{k_5}{\rightleftharpoons}} \tF,
\]
with forward fluxes as in \S6.1 and reverse fluxes
\[
\bar r_3 = k_{-3}\,\tS,\qquad \bar r_4 = k_{-4}\,\tP^4,\qquad \bar r_5 = k_{-5}\,\tF,\qquad k_{-i} = \epsilon\,\kappa_{-i}.
\]
Here $\epsilon$ is a single dimensionless scale controlling how far the network departs from strict irreversibility, and each $\kappa_{-i}$ carries the units its own reaction order requires---$S^{-1}$ for $\kappa_{-3}$ and $\kappa_{-5}$, $S^{-1}D^{-3}$ for $\kappa_{-4}$. A single dimensional parameter could not serve all three, since the reverse steps differ in order. In the non-dimensional calculations reported below we set $\kappa_{-i}=1$, so that numerically $\bar r_3 = \epsilon\tS$, $\bar r_4 = \epsilon\tP^4$, $\bar r_5 = \epsilon\tF$. The equilibrium constants are $K_3 = k_3/k_{-3}$, $K_4 = k_4/k_{-4}$, $K_5 = k_5/k_{-5}$, equal to $k_i/\epsilon$ under that normalisation. The balances are
\begin{gather*}
\dd\tF/\dd s = 2r_2 - 2r_1 + r_5 - \bar r_5,\qquad \dd\tN/\dd s = r_1 - r_2 - r_3 + \bar r_3,\\
\dd\tP/\dd s = 4r_4 - 4\bar r_4 - 2r_3 + 2\bar r_3 - r_5 + \bar r_5,\qquad \dd\tS/\dd s = r_3 - \bar r_3 - r_4 + \bar r_4.
\end{gather*}
Each reverse step carries the negated stoichiometry of its forward step, so the weight vector $(1,2,1,4)$ annihilates the reversible system as well and the budget is preserved exactly (verified to $5.7\times10^{-14}$ over 2000 random states at $\epsilon\in\{10^{-3},10^{-2},10^{-1}\}$).

Two structural facts change: \eqref{eq:factor2} no longer factors, since the reverse fluxes contribute terms independent of $\tP$; and $\tP=0$ is no longer invariant, since $\bar r_5$ produces past events from future ones. The past-poor state is replaced by a past-sparse state. In transition-state terms, the reverse step $\bar r_3$ is reversion of the activated complex to its reactants, and its competition with completion defines the forward-commitment probability \eqref{eq:transmission} of \S7.

Steady states were located by multistart root finding: 400 starts per leaf drawn uniformly and rescaled to the target budget, solved by a Powell hybrid method, retained when the residual of $(\dd\tF/\dd s, \dd\tN/\dd s, \dd\tS/\dd s, E-E_0)$ fell below $10^{-10}$ with all components non-negative, and merged when agreeing to atol $=10^{-6}$, rtol $=10^{-4}$. Because multistart search alone cannot prove that no steady state has been missed, the count was confirmed by a one-dimensional parameterisation, written here with unit forward coefficients $k_1=k_2=k_3=k_4=1$ and $\kappa_{-i}=1$, as in the calculations reported. At a steady state the net fluxes satisfy $r_1-r_2 = r_3-\bar r_3 = r_4-\bar r_4 = J$ and $r_5-\bar r_5 = 2J$, so for given $\tP>0$ one has $J=(k_5\tP-\epsilon\tF)/2$, $\tS = J+\epsilon\tP^4$ and $\tN = (J+\epsilon\tS)/\tP^2$, and $\tF^2 = \tN + J$ becomes
\[
\tF^2 + \tfrac{c\,\epsilon}{2}\,\tF - \Bigl(\tfrac{c\,k_5\tP}{2} + \epsilon^2\tP^2\Bigr) = 0,\qquad c = 1 + \frac{1+\epsilon}{\tP^2}.
\]
This quadratic has a positive linear coefficient and a negative constant term, so for every $\tP>0$ it has exactly one positive root. Each $\tP$ therefore determines at most one steady-state composition, and the steady states on a leaf are exactly the crossings of the single curve $E(\tP)$ with the budget. That uniqueness is analytical; locating the crossings is numerical, and the claim is confined to the scanned range $10^{-7}\le\tP\le9$. Above it $E(\tP)$ increases monotonically and exceeds every budget considered; below it no claim is made. At $E=9$ the scan returns three crossings at each of $\epsilon=10^{-2},10^{-3},10^{-4}$---at $\epsilon=10^{-3}$, $\tP=0.001912$, $0.147298$, $2.420537$---in agreement with the multistart results. Fold budgets were bisected on the retained-root count to 26 iterations; fold positions in $k_5$ were found by quasi-static continuation in steps of 0.02, recording the first step at which the branch was lost.

\paragraph{The fold survives.} Multistart root finding at $\epsilon=10^{-3}$ returns three states per leaf above threshold ($\tP = 0.0019$, $0.1473$, $2.4205$ at $E=9$; outer two stable, middle a saddle), and the fold budget converges to its irreversible value: 4.5292, 4.5263, 4.4966, 4.4252 at $\epsilon = 10^{-4}$, $10^{-3}$, $10^{-2}$, $3\times10^{-2}$, against 4.5295 analytically.

\begin{figure}[t]
\centering
\includegraphics[width=\textwidth]{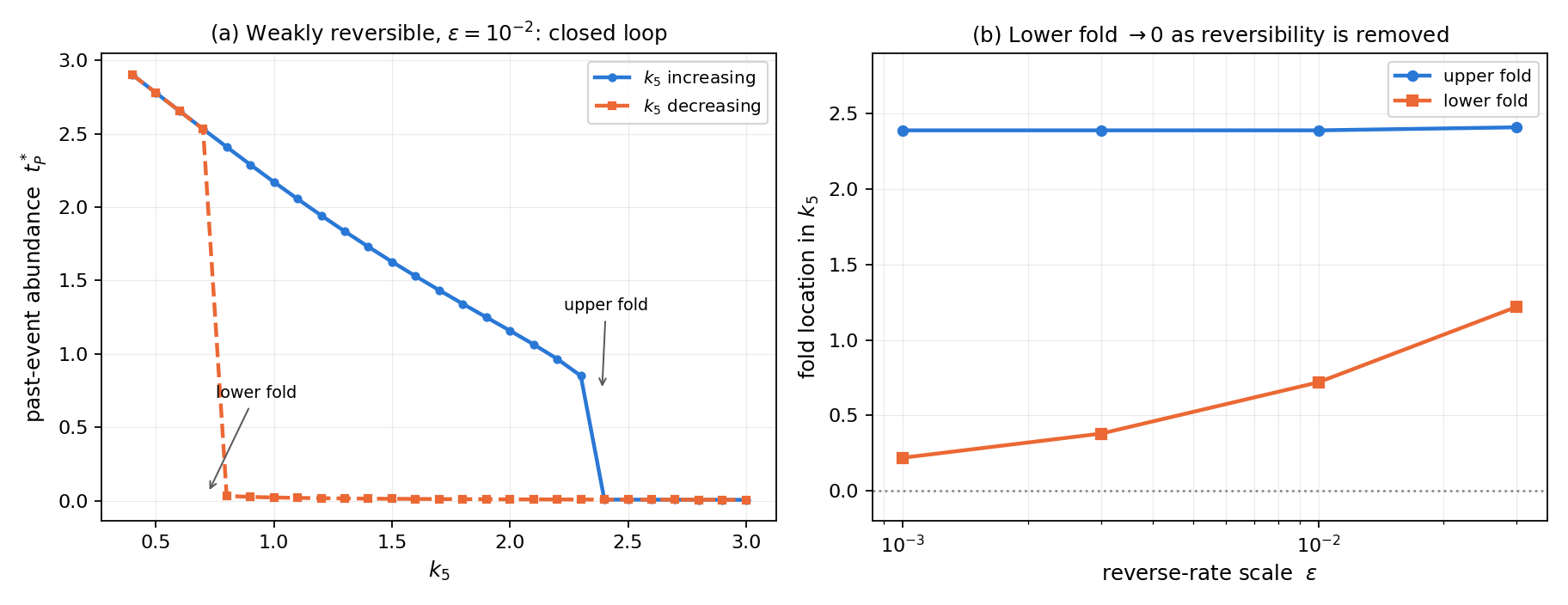}
\caption{Weakly reversible network at $E=9$. (a) Up-and-down $k_5$ sweep at $\epsilon=10^{-2}$ closes into a two-fold loop. (b) The upper fold is insensitive to $\epsilon$; the lower fold is driven toward $k_5=0$ as reversibility vanishes.}
\label{fig:loop}
\end{figure}

Because the past-sparse branch can itself lose stability, the $k_5$ sweep closes into a two-fold hysteresis loop (Fig.~\ref{fig:loop}a). The upper fold is insensitive to $\epsilon$, remaining near $k_5=2.39$, while the lower fold approaches $k_5=0$ as reversibility is removed (Fig.~\ref{fig:loop}b). One-way latching is thus the singular limit of ordinary hysteresis, with the recovery threshold driven to the boundary of parameter space.

The scaling can be obtained analytically. Near the past-sparse state $\tP$ is small, so $\tS\simeq k_3\tN\tP^2/k_4$ from the $\tS$ balance, and the $\tP$ balance reduces at leading order to
\[
4k_4\tS - 2k_3\tN\tP^2 - k_5\tP + \epsilon\kappa_{-5}\tF \simeq 2k_3\tN\tP^2 - k_5\tP + \epsilon\kappa_{-5}\tF = 0,
\]
a quadratic in $\tP$ whose two roots are the past-sparse state and the saddle. They merge---which is the lower fold---when its discriminant vanishes:
\begin{equation}
k_{5,\mathrm{lower}} \simeq \bigl(8\,k_3\,\tN\,\tF\,\epsilon\,\kappa_{-5}\bigr)^{1/2},\label{eq:lower}
\end{equation}
with $\tF$ and $\tN$ evaluated at the past-poor state, where they are fixed by $E$ alone through $k_1\tF^2 = k_2\tN$ and $E=\tF+2\tN$. The law is therefore $\propto\sqrt\epsilon$ with a coefficient predicted from the budget. At unit constants \eqref{eq:lower} gives 5.196, 7.326 and 11.180 at $E=6,9,15$.

Bisection on the switching parameter, to 32 iterations, confirms this: the measured coefficients are 5.174, 7.307, 11.166 at $\epsilon=10^{-3}$ and 5.173, 7.310, 11.167 at $\epsilon=10^{-4}$, agreeing with \eqref{eq:lower} to between 0.1\% and 0.4\%, with the residual shrinking as $\epsilon$ decreases, as expected of a leading-order asymptotic. Across a wider sweep at $E=9$:

\begin{table}[H]
\centering\small
\begin{tabular}{lrr}
\toprule
$\epsilon$ & $k_{5,\mathrm{lower}}$ & $k_{5,\mathrm{lower}}/\sqrt\epsilon$\\
\midrule
$3\times10^{-2}$ & 1.22727 & 7.086\\
$1\times10^{-2}$ & 0.72284 & 7.228\\
$3\times10^{-3}$ & 0.39903 & 7.285\\
$1\times10^{-3}$ & 0.23079 & 7.298\\
$3\times10^{-4}$ & 0.12657 & 7.308\\
$1\times10^{-4}$ & 0.07276 & 7.276\\
\bottomrule
\end{tabular}
\end{table}

A free power-law fit over these two and a half decades gives an exponent of 0.496, consistent with the derived value of $1/2$; the loop width is correspondingly $2.39 - (8k_3\tN\tF\epsilon)^{1/2}$, saturating only as $\epsilon\to0$.

Equation \eqref{eq:lower} makes the distinction of \S6.5 quantitative. The loop width measures how far the progression falls short of being recoverable by parameter reversal, and the degree of recoverability varies continuously with the reverse-rate scale in this family of models. The square-root law means the approach to strict one-wayness is slow: suppressing the reverse fluxes by two orders of magnitude reduces the residual recoverability by only one.

\section{Detailed balance excludes the fold}

Write $N$ for the stoichiometric matrix, species by reactions. Its \emph{left} null space, $\mathbf{w}^\top N = 0$, contains the conserved event-weight vector $\mathbf{w} = (1,2,1,4)$ of \eqref{eq:budget}. Its \emph{right} null space, $N\mathbf{v}=0$, contains the reaction cycles, and is one-dimensional here, generated by the flux vector $(\text{R1},\text{R3}',\text{R4}',\text{R5}) = (1,1,1,2)$:
\[
(2\tF\to\tN) + (\tN+2\tP\to\tS) + (\tS\to4\tP) + 2\,(\tP\to\tF) = 0.
\]
Detailed balance imposes exactly one Wegscheider condition, $K_1K_3K_4K_5^2 = 1$. The parameterisation of Appendix~B, with $\kappa_{-i}=1$ and $K_i = k_i/\epsilon$, gives $\epsilon^{-4}$, which equals unity only at $\epsilon=1$, where the rate constants satisfy the detailed-balance condition and the network admits an equilibrium at which every forward reaction is balanced by its reverse; away from that value the network is reversible but not detailed balanced. Imposing the condition by setting $K_5 = (K_1K_3K_4)^{-1/2}$ and repeating the multistart search returns one steady state per leaf at every budget tested ($E=4,9,20$ at $K_3=K_4=10,100,1000$).

The reason is classical. A detailed-balanced mass-action system admits a strictly convex free-energy Lyapunov function and possesses a unique positive equilibrium in each positive stoichiometric compatibility class; under the standard detailed-balance results applicable here, trajectories with strictly positive initial conditions converge to that equilibrium~\cite{hornjackson1972,feinberg1987}. We make no claim about boundary trajectories or about persistence, neither of which is needed for the conclusion: multistationarity in the positive class is excluded, so no fold can exist there. Direct verification at $K_3=K_4=100$, $E=9$: every step individually balanced to $\le10^{-11}$ at the equilibrium, and the free energy decreasing monotonically along trajectories launched from $\tP = 0.9$, $1.0$, $6.6$, all converging to that state.

The bistability of \S6 is therefore not equilibrium bistability. Note that what fails here is detailed balance, not reversibility: Appendix~B exhibits a fully reversible network that remains bistable, and the fold is lost only once the Wegscheider condition is enforced. In the chemically interpreted analogue the requirement is that the cycle carry nonzero affinity. Under the energy reading of \S4.4 the event model assigns an energy content to each class but posits no temperature or entropy function, so the affinity condition is not derived thermodynamically; within the event model it maps onto the kinetic asymmetry of the progression, which remains a premise, not a derived result. Section~9 gives the relational reading of how that asymmetry is sustained in a closed network.

\section*{Data and code availability}

All results in this manuscript are generated by the numerical procedures described in Appendix~A. The integration, continuation, root-finding and figure-generation scripts are archived on Zenodo at \url{https://doi.org/10.5281/zenodo.21815545}.

\section*{Competing interests}

The author declares no competing interests.

\section*{Use of AI assistance}

The work reported here was carried out with the assistance of Claude, a large language model developed by Anthropic. Because that assistance extended well beyond language editing, its scope is set out in full.

The assistant performed the numerical and symbolic computation: the steady-state and stability analysis, the Radau integrations, the multistart root finding for the reversible and detailed-balanced networks, the control-coefficient calculations, and the null-space and Wegscheider algebra. It generated all figures. It contributed several of the results reported above, among them the factorisation of the first-order steady-state condition for arbitrary completion yield, the condition $k_2=k_5$ under which the reduced branch equation is reciprocally symmetric, and the asymptotic scaling of the lower switching threshold together with its derivation from the discriminant of the near-boundary balance. It also identified the stoichiometric inconsistency in the earlier formulation of the network and the obstruction to folding under first-order participation. It drafted the manuscript text, including the revisions of versions~2 and~3, and carried out the independent re-implementation reported in Appendix~A. That re-implementation was performed by the same AI system that produced the original calculations, using different code and methods; it is therefore a check against implementation error, not an independent review by a separate party. The error in the constant of \eqref{eq:asym} was identified by an external critique of the August 2026 draft.

The author conceived the event-progression model and the relational framework on which it rests, directed the line of inquiry, and made the interpretive and scope decisions---the treatment of the system as closed, the standing of the boundary configuration under relational labelling, the definition of the progression parameter, the deferral of spatial extension, and, in version~2, the reading of events as discrete parcels of energy, the identification of the activated class with the transition state and the present, and the statistical reading of the rate constants; and, in version~3, the heat analogy, the relational reading of renewal and of entropy, and the decision not to invoke a low-entropy initial condition. The author verified all references and takes full responsibility for the content of this paper, including the correctness of its results and the accuracy of its citations. No AI system is an author of this work.

\end{document}